\documentclass[fleqn,12pt]{article}

\usepackage{latexsym,ifthen,epsfig,color}
\usepackage{amsmath,amssymb,amsthm}
\usepackage{float}

\newcommand{\join}{\text{\textcircled{{\footnotesize 1}}}}
\newcommand{\cojoin}{\text{\textcircled{{\footnotesize 0}}}}

\newcommand{\NP}{\ensuremath{\mathbb{NP}}}

\newtheorem{clai}{Claim}[section]
\newtheorem{theo}{Theorem}
\newtheorem{lemma}{Lemma}

\newtheorem{coro}{Corollary}

\begin{document}

\author{
Andreas Brandst\"adt\\
\small Institut f\"ur Informatik, Universit\"at Rostock, D-18051 Rostock, Germany\\
\small \texttt{andreas.brandstaedt@uni-rostock.de}\\
\and
Raffaele Mosca\\
\small Dipartimento di Economia, Universit\'a degli Studi ``G. D'Annunzio'',
Pescara 65121, Italy\\
\small \texttt{r.mosca@unich.it}
}

\title{Finding Efficient Domination for $P_8$-Free Bipartite Graphs in Polynomial Time}

\maketitle

\begin{abstract}
A vertex set $D$ in a finite undirected graph $G$ is an {\em efficient dominating set} (\emph{e.d.s.}\ for short) of $G$ if every vertex of $G$ is dominated by exactly one vertex of $D$. The \emph{Efficient Domination} (ED) problem, which asks for the existence of an e.d.s.\ in $G$, is known to be \NP-complete for $P_7$-free graphs, and even for very restricted $H$-free bipartite graph classes such as for $K_{1,4}$-free bipartite graphs as well as for $C_4$-free bipartite graphs while it is solvable in polynomial time for $P_7$-free bipartite graphs as well as for $S_{2,2,4}$-free bipartite graphs. Here we show that ED can be solved in polynomial time for $P_8$-free bipartite graphs. 
\end{abstract}


\section{Introduction}\label{sec:intro}

Let $G=(V,E)$ be a finite undirected graph. A vertex $v$ {\em dominates} itself and its neighbors. A vertex subset $D \subseteq V$ is an {\em efficient dominating set} ({\em e.d.s.}\ for short) of $G$ if every vertex of $G$ is dominated by exactly one vertex in $D$; for any e.d.s.\ $D$ of $G$, $|D \cap N[v]| = 1$ for every $v \in V$ (where $N[v]$ denotes the closed neighborhood of $v$).
Note that not every graph has an e.d.s.; the {\sc Efficient Dominating Set} ({\sc ED} for short) problem asks for the existence of an e.d.s.\ in a given graph~$G$.
The notion of efficient domination was introduced by Biggs \cite{Biggs1973} under the name {\em perfect code}.

\medskip

The Exact Cover Problem asks for a subset ${\cal F'}$ of a set family ${\cal F}$ over a ground set, say $V$, containing every vertex in $V$ exactly once, i.e., ${\cal F'}$ forms a partition of $V$. As shown by Karp \cite{Karp1972}, this problem is \NP-complete even for set families containing only $3$-element subsets of $V$ (see problem X3C [SP2] in \cite{GarJoh1979}).

\medskip

Clearly, ED is the Exact Cover problem for the closed neighborhood hypergraph of $G$, i.e., if $D=\{d_1,\ldots,d_k\}$ is an e.d.s.\ of $G$ then $N[d_1] \cup \ldots \cup N[d_k]$ forms a partition of $V$ (we call it the {\em e.d.s.\ property}). In particular, the distance between any pair of distinct $D$-vertices $d_i,d_j$ is at least 3. 
In \cite{BanBarSla1988,BanBarHosSla1996}, it was shown that the ED problem is \NP-complete. 

\medskip

For a set ${\cal F}$ of graphs, a graph $G$ is called {\em ${\cal F}$-free} if $G$ contains no induced subgraph isomorphic to a member of ${\cal F}$; in particular, we say that $G$ is {\em $H$-free} if $G$ is $\{H\}$-free. Let $H_1+H_2$ denote the disjoint union of graphs $H_1$ and $H_2$, and for $k \ge 2$, let $kH$ denote the disjoint union of $k$ copies of $H$.
For $i \ge 1$, let $P_i$ denote the chordless path with $i$ vertices, and let $K_i$ denote the complete graph with $i$ vertices (clearly, $P_2=K_2$).
For $i \ge 4$, let $C_i$ denote the chordless cycle with $i$ vertices.

For indices $i,j,k \ge 0$, let $S_{i,j,k}$ denote the graph with vertices $u,x_1,\ldots,x_i$, $y_1,\ldots,y_j$, $z_1,\ldots,z_k$ (and {\em center} $u$) such that the subgraph induced by $u,x_1,\ldots,x_i$ forms a $P_{i+1}$ ($u,x_1,\ldots,x_i$), the subgraph induced by $u,y_1,\ldots,y_j$ forms a $P_{j+1}$ ($u,y_1,\ldots,y_j$), and the subgraph induced by $u,z_1,\ldots,z_k$ forms a $P_{k+1}$ ($u,z_1,\ldots,z_k$), and there are no other edges in $S_{i,j,k}$. Thus, claw is $S_{1,1,1}$, chair is $S_{1,1,2}$, and $P_k$ is isomorphic to $S_{0,0,k-1}$. 

\medskip

In \cite{EscWan2014,SmaSla1995,YenLee1996}, it was shown that ED is \NP-complete for $2P_3$-free chordal unipolar graphs and thus, in general, for $P_7$-free graphs.
In \cite{BraMos2016}, ED is solvable in polynomial time for $P_6$-free graphs (which leads to a dichotomy). 

\medskip

A bipartite graph $G$ is {\em chordal bipartite} if $G$ is $C_{2k}$-free for every $k \ge 3$. 
Lu and Tang \cite{LuTan2002} showed that ED is \NP-complete for chordal bipartite graphs (i.e., hole-free bipartite graphs). Thus, for every $k \ge 3$, ED is \NP-complete for $C_{2k}$-free bipartite graphs.
Moreover, ED is \NP-complete for planar bipartite graphs \cite{LuTan2002} and even for planar bipartite graphs of maximum degree 3 \cite{BraMilNev2013} and girth at least $g$ for every fixed $g$ \cite{Nevri2014}. Thus, ED is \NP-complete for $K_{1,4}$-free bipartite graphs and for $C_4$-free bipartite graphs.

\medskip

In \cite{BraFicLeiMil2015}, it is shown that one can extend polynomial time algorithms for Efficient Domination to such algorithms for weighted Efficient Domination. Thus, from now on, we focus on the unweighted ED problem.

\medskip

In \cite{BraLeiRau2012}, it is shown that ED is solvable in polynomial time for AT-free graphs. Moreover,
in \cite{BraLeiRau2012}, it is shown that ED is solvable in polynomial time for interval bigraphs, and convex bipartite graphs are a subclass of them (and of chordal 
bipartite graphs). Moreover, Lu and Tang \cite{LuTan2002} showed that ED is solvable in linear time for bipartite permutation graphs (which is a subclass of convex bipartite graphs). 

\begin{figure}[ht]
  \begin{center}
   \epsfig{file=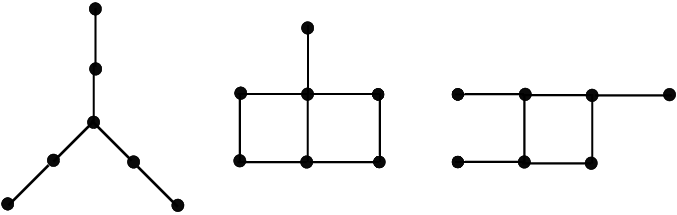}
   \caption{Forbidden induced subgraphs $H_1=S_{2,2,2},H_2,H_3$ for bipartite permutation graphs.}
   \label{S222etc}
  \end{center}
\end{figure}

It is well known (see e.g.\ \cite{BraLeSpi1999,Koehl1999}) that $G$ is a bipartite permutation graph if and only if $G$ is AT-free bipartite if and only if $G$ is 
($H_1,H_2,H_3$,hole)-free bipartite (see Figure \ref{S222etc}).
Thus, while ED is \NP-complete for $(H_2,H_3)$-free bipartite graphs (since $H_2$ and $H_3$ contain $C_4$ and $H_2$ contains $K_{1,4}$), in \cite{BraMos2019} we have shown that ED is solvable in polynomial time for $S_{2,2,2}$-free (and more generally, for $S_{2,2,4}$-free) bipartite graphs.

\medskip

In \cite{BraMos2019} we have shown that ED is solvable in polynomial time for $P_7$-free bipartite graphs as well as for $S_{2,2,4}$-free bipartite graphs. Now in this manuscript, we show: 

\begin{theo}\label{EDP8frbippol}
For $P_8$-free bipartite graphs, the ED problem is solvable in polynomial time. 
\end{theo} 

\section{Preliminaries}

Recall that $G=(X,Y,E)$ is a $P_8$-free bipartite graph, i.e., every vertex in $X$ is black, every vertex in $Y$ is white, and $D$ is a possible e.d.s.\ of $G$. 

\medskip

A vertex $u$ {\em contacts} $v$ if $uv \in E$. For a subset $U \subseteq X \cup Y$, a vertex $v \notin U$ {\em contacts $U$} if $v$ has a neighbor in $U$. 
Moreover, a vertex $v \notin U$ has a {\em join} to $U$, say $v \join U$, if $v$ contacts every vertex in $U$,
and $v$ has a {\em co-join} to $U$, say $v \cojoin U$, if $v$ does not contact any vertex in $U$. 

\medskip

A vertex $u \in V$ is {\em forced} if $u \in D$ for every e.d.s.\ $D$ of $G$; $u$ is {\em excluded} if $u \notin D$ for every e.d.s.\ $D$ of $G$.
For example, if $x_1,x_2 \in X$ are leaves in $G$ and $y$ is the neighbor of $x_1,x_2$ then $x_1,x_2$ are excluded and $y$ is forced.

\medskip

By a forced vertex, $G$ can be reduced to $G'$ as follows:

\begin{clai}\label{forcedreduction}
If $u$ is forced then $G$ has an e.d.s.\ $D$ with $u \in D$ if and only if the reduced graph $G'=G \setminus N[u]$ has an e.d.s.\ $D'=D \setminus \{u\}$ such that all vertices in $N^2(u)$ are excluded in $G'$.
\end{clai}

Analogously, if we assume that $v \in D$ for a vertex $v \in V=X \cup Y$ then $u \in V$ is {\em $v$-forced} if $u \in D$ for every e.d.s.\ $D$ of $G$ with $v \in D$,
and $u$ is {\em $v$-excluded} if $u \notin D$ for every e.d.s.\ $D$ of $G$ with $v \in D$. For checking whether $G$ has an e.d.s.\ $D$ with $v \in D$, we can clearly reduce $G$ by forced vertices as well as by $v$-forced vertices when we assume that $v \in D$:

\begin{clai}\label{vforcedreduction}
If we assume that $v \in D$ and $u$ is $v$-forced then $G$ has an e.d.s.\ $D$ with $v \in D$ if and only if the reduced graph $G'=G \setminus N[u]$ has an e.d.s.\ $D'=D \setminus \{u\}$ with $v \in D'$ such that all vertices in $N^2(u)$ are $v$-excluded in $G'$.
\end{clai}

Similarly, for $k \ge 2$, $u \in V$ is {\em $(v_1,\ldots,v_k)$-forced} if $u \in D$ for every e.d.s.\ $D$ of $G$ with $v_1,\ldots,v_k \in D$, and correspondingly, $u \in V$ is {\em $(v_1,\ldots,v_k)$-excluded} if $u \notin D$ for such e.d.s.\ $D$, and $G$ can be reduced by the same principle.

\medskip

Clearly, the e.d.s.\ problem in $G=(X,Y,E)$ can be done independently for every connected component in $G$. 
Thus, we can assume that $G$ is connected.   

\medskip

Let $dist_G(u,v)$ denote the minimum distance, i.e., the minimum length of a path between $u$ and $v$ in $G$. 
By the e.d.s.\ property, the distance between two $D$-vertices is at least 3. Moreover, we have:

\begin{clai}\label{distDXY}
If the $P_8$-free bipartite graph $G=(X,Y,E)$ is connected then:
\begin{itemize}
\item[$(i)$] For every $x \in D \cap X$ and $y \in D \cap Y$, $dist_G(x,y)=3$ or $dist_G(x,y)=5$. 
\item[$(ii)$] For every $v,v' \in D \cap X$ or $v,v' \in D \cap Y$, $dist_G(v,v')=4$ or $dist_G(v,v')=6$. 
\end{itemize}
\end{clai}

\noindent
{\bf Proof.}
$(i)$: By the e.d.s.\ property and since $G$ is bipartite, for $x \in D \cap X$ and $y \in D \cap Y$, $dist_G(x,y) \ge 3$. 
In particular, $dist_G(x,y) \ge 2k+1$, $k \ge 1$. 

If $dist_G(x,y) = 7$ then $(x,y_1,x_1,y_2,x_2,y_3,x_3,y)$ induce a $P_8$ in $G$, which is a contradiction. Thus, $dist_G(x,y)=3$ or $dist_G(x,y)=5$.

\medskip

\noindent
$(ii)$: Without loss of generality, assume that $v,v' \in D \cap X$. Then by the e.d.s.\ property and since $G$ is bipartite, $dist_G(v,v') \ge 4$.  
In particular, $dist_G(x,y) \ge 2k$, $k \ge 2$. If If $dist_G(v,v') = 8$ then there is a $P_8$ in $G$, which is a contradiction. Thus, $dist_G(v,v')=4$ or $dist_G(v,v')=6$. Analogously, if $v,v' \in D \cap Y$ then $dist_G(v,v')=4$ or $dist_G(v,v')=6$.

\medskip

Thus, Claim \ref{distDXY} is shown.
\qed

\medskip

Recall that $G$ is connected. If for an e.d.s.\ $D$ in the $P_8$-free bipartite graph $G=(X,Y,E)$, $|D| = 1$, say without loss of generality, $D=\{x\}$, then 
$x \join Y$ and $X=\{x\}$ which is trivial. Now assume that $|D| \ge 2$ and without loss of generality, $D \cap X \neq \emptyset$. 

If $D \cap Y = \emptyset$ and $x_1,x_2 \in D$ then 
by the e.d.s.\ property, $x_1$ and $x_2$ do not have any common neighbors in $Y$, i.e., $N(x_1) \cap N(x_2)=\emptyset$, and if $y_1x \in E$, $y_2x \in E$ then 
by the e.d.s.\ property, $x \notin D$ and $x$ must have a $D$-neighbor in $Y$ but $D \cap Y = \emptyset$. 
Thus, for every $y_i \in N(x_i)$, $i \in {1,2}$, $y_1$ and $y_2$ do not have any common neighbor in $X$, i.e., $N[x_1]$ as well as $N[x_2]$ are disconnected, which is a contradiction since $G$ is connected. Thus, $D \cap X \neq \emptyset$ and $D \cap Y \neq \emptyset$, i.e., $|D \cap X| \ge 1$ and $|D \cap Y| \ge 1$. 

\medskip

Recall that in \cite{BraMos2019}, ED is solvable in polynomial time for $P_7$-free bipartite graphs. Now assume that there are $P_7$'s in $G$. 

\medskip

First assume that $|D \cap X| = 1$, say $D \cap X = \{x\}$. Then every $y \in Y \setminus N(x)$ does not have any $D$-neighbor, and $D=\{x\} \cup (Y \setminus N(x))$ or there is no e.d.s.\ in $G$ if for $y,y' \in Y \setminus N(x)$, $dist_G(y,y')=2$, which is a trivial e.d.s.\ solution. 
Analogously, if $|D \cap Y| = 1$ then it is a trivial e.d.s.\ solution. 

From now on, assume that $|D \cap X| \ge 2$ and $|D \cap Y| \ge 2$.

\medskip

Recall that by Claim \ref{distDXY}, for every $x \in D \cap X$ and $y \in D \cap Y$, $dist_G(x,y)=3$ or $dist_G(x,y)=5$, and for every $v,v' \in D \cap X$ or 
$v,v' \in D \cap Y$, $dist_G(v,v')=4$ or $dist_G(v,v')=6$. 

\medskip

Assume that $x \in D \cap X$ with $P_3$ $(x,y_1,x_1)$ in $G$. Then by the e.d.s.\ property, $x_1$ must have a $D$-neighbor $y \in D \cap Y$, i.e., $(x,y_1,x_1,y)$ induce a $P_4$ in $G$, and $dist_G(x,y)=3$.  

\medskip

If $|N(x)|=1$ and $|N(y)|=1$, say $(x,y_1,x_1,y)$ induce a $P_4$ with leaves $x,y$, then $x_1,y_1$ are excluded, $x,y$ are forced, and $G$ can be reduced as in 
Claim \ref{forcedreduction}. 
Thus, from now on, assume that either $|N(x)| \ge 2$ or $|N(y)| \ge 2$ (and there are no leaves in $G$ with distance 3); 
let $N(x)=\{y_1,\ldots,y_i\}$ and $N(y)=\{x_1,\ldots,x_j\}$, $i \ge 2$ or $j \ge 2$, and let $x_1y_1 \in E$ 
(possibly there could be more edges between $N(x)$ and $N(y)$).     

\medskip

Let $N_0:=D_{basis}$, i.e., $x,y \in D_{basis}$ with $dist_G(x,y) = 3$, say $(x,y_1,x_1,y)$ induce a $P_4$ in $G$, and every $(x,y)$-forced vertex is in $D_{basis}$. 
Moreover, let $N_i$, $i \ge 1$, be the distance levels of $D_{basis}$. 
Since $G$ is $P_8$-free bipartite, we have $N_7=\emptyset$ (else $(r_0,r_1,\ldots,r_7)$ with $r_i \in N_i$, $0 \le i \le 7$, induce a $P_8$ in $G$). 
By the e.d.s.\ property, we have: 
\begin{equation}\label{DN1N2empty}
D \cap (N_1 \cup N_2)=\emptyset. 
\end{equation}
If there is a $u \in N_2$ with $N(u) \cap N_3=\emptyset$ then by the e.d.s.\ property, there is no such e.d.s.\ $D$ in $G$ with $N_0=D_{basis}$. Thus, we assume:
\begin{equation}\label{uinN2neighbinN3}
\mbox{ Every vertex } u \in N_2 \mbox{ has a $D$-neighbor in } N_3.
\end{equation}
If for $u \in N_2$, $|N(u) \cap N_3|=1$, say $N(u) \cap N_3=\{v\}$, then by (\ref{DN1N2empty}) and the e.d.s.\ property, $v$ is $D_{basis}$-forced. 
Moreover, if there are $u,u' \in N_2$ with $N(u) \cap N_3=\{v\}$, $N(u') \cap N_3=\{v'\}$ and $dist_G(v,v')=1$ or $dist_G(v,v')=2$ then by the e.d.s.\ property, there is no such e.d.s.\ $D$ in $G$ with $N_0=D_{basis}$. Thus assume that for $N(u) \cap N_3=\{v\}$, $N(u') \cap N_3=\{v'\}$, $dist_G(v,v') \ge 3$.  

Then one can update $D_{basis}:= D_{basis} \cup \{v\}$ and redefine the distance levels $N_i$, $i \ge 1$, with respect to $D_{basis}$. 
Thus, we assume:   
\begin{equation}\label{uinN2twoneighbinN3}
\mbox{ For every } u \in N_2, |N(u) \cap N_3| \ge 2.
\end{equation}
If for $v \in N_3$, $N(v) \cap (N_3 \cup N_4)=\emptyset$ then by (\ref{DN1N2empty}) and the e.d.s.\ property, $v$ is $D_{basis}$-forced. 
Moreover, if there are two such $v_i \in N_3$, $N(v_i) \cap (N_3 \cup N_4)=\emptyset$, $i \in \{1,2\}$, with common neighbor $u \in N_2$, i.e., 
$uv_i \in E$, $i \in \{1,2\}$, then by the e.d.s.\ property, there is no such e.d.s.\ $D$ in $G$ with $N_0=D_{basis}$. 
Thus assume that such $D_{basis}$-forced vertices $v_i \in N_3$, $i \in \{1,2\}$, do not have any common neighbor in $N_2$. 

Then for every $v\in N_3$ with $N(v) \cap (N_3 \cup N_4)=\emptyset$, one can update $D_{basis}:= D_{basis} \cup \{v\}$ and redefine the distance levels 
$N_i$, $i \ge 1$, with respect to $D_{basis}$ as above. 
Thus, we assume:
\begin{equation}\label{vinN3neighbinN3N4}
\mbox{ Every } v \in N_3 \mbox{ has a neighbor in } N_3 \cup N_4.
\end{equation}

\begin{clai}\label{uinN2noendpointP6inN0N1}
Every $u \in N_2$ is no endpoint of a $P_k$ $P=(w_1,\ldots,w_{k-1},u)$, $k \ge 6$, with $w_1,\ldots,w_{k-2} \in N_0 \cup N_1$ and $w_{k-1} \in N_1$.
\end{clai}

\noindent
{\bf Proof.}
Let $u \in N_2$. Since $G$ is $P_8$-free bipartite, $u$ is no endpoint of any $P_8$ $P=(w_1,\ldots,w_7,u)$ with $w_1,\ldots,w_6 \in N_0 \cup N_1$ and $w_7 \in N_1$.
 
\medskip

If $u$ is the endpoint of a $P_7$ $P=(w_1,\ldots,w_6,u)$ with $w_1,\ldots,w_5 \in N_0 \cup N_1$ and $w_6 \in N_1$ then, since by (\ref{DN1N2empty}), 
$D \cap (N_1 \cup N_2)=\emptyset$, $u$ has no $D$-neighbor (else there is a $P_8$ $P=(w_1,\ldots,w_6,u,v)$ with $D$-neighbor $v \in D \cap N_3$ of $u$ in $G$) and there is no such e.d.s. Thus, $u$ is the endpoint of a $P_k$, $k \le 6$, $P=(w_1,\ldots,w_{k-1},u)$ with $w_1,\ldots,w_{k-2} \in N_0 \cup N_1$ and $w_{k-1} \in N_1$. 

\medskip

If $u \in N_2$ is the endpoint of a $P_6$ $P=(w_1,\ldots,w_5,u)$ with $w_1,\ldots,w_4 \in N_0 \cup N_1$ and $w_5 \in N_1$ then $u$ must have a 
$D$-neighbor $v \in D \cap N_3$, and since $G$ is $P_8$-free, $v$ has no further neighbor $w \in N_3 \cup N_4$ (else there is a $P_8$ $(w_1,\ldots,w_5,u,v,w)$ in $G$). Thus, $v$ is $D_{basis}$-forced, i.e., $D_{basis}:=D_{basis} \cup \{v\}$ and $D_{basis}$ (and its distance levels) can be updated. 

\medskip

Moreover, if $u$ has two such neighbors $v,v' \in N_3$ then assume that $v' \notin D \cap N_3$, and $v'$ must have a $D$-neighbor $w' \in D \cap (N_3 \cup N_4)$. But then $(w_1,\ldots,w_5,u,v',w')$ induce a $P_8$ in $G$, which is a contradiction. Thus, there is no such e.d.s.\ with $N_0=D_{basis}$.

\medskip

Thus, $u$ is no endpoint of such a $P_k$ $P=(w_1,\ldots,w_{k-1},u)$, $k \ge 6$, and Claim \ref{uinN2noendpointP6inN0N1} is shown.
\qed

\medskip

Recall that $x,y \in D_{basis}$ and by every $(x,y)$-forced vertex $v \in D$, $D_{basis}$ was updated, i.e., $D_{basis}:=D_{basis} \cup \{v\}$.

\begin{lemma}\label{N2endpointP4}
For $N_0=D_{basis}$ with $x \in N_0 \cap X$, $y \in N_0 \cap Y$, we have:
\begin{itemize}
\item[$(i)$] If $u \in N_2 \cap X$ and $u \cojoin N(v)$ for some $v \in N_0 \cap X$ (possibly $v=x$), or $u \in N_2 \cap Y$ and $u \cojoin N(v)$ for some 
$v \in N_0 \cap Y$ (possibly $v=y$), then $u$ is the endpoint of a $P_5$ whose remaining vertices are in $N_0 \cup N_1$.
\item[$(ii)$] If $u \in N_2 \cap X$ contacts all $N(v)$ with $v \in N_0 \cap X$ (also $v=x$), or $u \in N_2 \cap Y$ contacts all $N(v)$ with $v \in N_0 \cap Y$ 
(also $v=y$) then $u$ is the endpoint of a $P_4$ whose remaining vertices are in $N_0 \cup N_1$. 
\item[$(iii)$] $N_6=\emptyset$, $N_5$ is independent, and every edge in $N_4$ does not contact $N_5$.
\end{itemize}
\end{lemma}

\noindent
{\bf Proof.}
$(i)$: Without loss of generality, assume that $u \in N_2 \cap X$. Clearly, for every $v \in N_0 \cap Y$ (possibly $v=y$), $u \cojoin N(v)$ since $N(v) \subset X$.

\medskip

\noindent
{\em Case $1$.} $u \cojoin N(x)$: 

\medskip

Since $u \in N_2$, $u$ must have a neighbor in $N_1$, say $uw \in E$ for some $w \in N(v) \cap N_1$, $v \in N_0 \cap X$, $v \neq x$. 
Since $v \in N_0 \cap X$ was $y$-forced, we have $dist_G(v,y)=3$, say $(v,y',x',y)$ induce a $P_4$ in $G[N_0 \cup N_1]$ with $x',y' \in N_1$. 

First assume that $uy' \in E$. If $y'x_1 \in E$ then $(u,y',x_1,y_1,x)$ induce a $P_5$ with endpoint $u$. Moreover, 
if $y'x_1 \notin E$ and $x'y_1 \in E$ then $(u,y',x',y_1,x)$ induce a $P_5$ with endpoint $u$. 
Now assume that $y'x_1 \notin E$ and $x'y_1 \notin E$. But then $(u,y',x',y,x_1,y_1,x)$ induce a $P_7$ with endpoint $u$, which is impossible 
by Claim \ref{uinN2noendpointP6inN0N1}.

Next assume that $uy' \notin E$, say $uw \in E$, $w \neq y'$, $w \in N(v)$. Since by Claim \ref{uinN2noendpointP6inN0N1}, $(u,w,v,y',x',y)$ does not induce a $P_6$ in $G$, we have $wx' \in E$. Again, if $wx_1 \in E$ then $(u,w,x_1,y_1,x)$ induce a $P_5$ with endpoint $u$, and if $wx_1 \notin E$ but $x'y_1 \in E$ then $(u,w,x',y_1,x)$ induce a $P_5$ with endpoint $u$. Finally, assume that $wx_1 \notin E$ and $x'y_1 \notin E$. But then again $(u,w,x',y,x_1,y_1,x)$ induce a $P_7$ with endpoint $u$, which is impossible by Claim \ref{uinN2noendpointP6inN0N1}. 

\medskip

\noindent
{\em Case $2$.} $u$ contacts $N(x)$ but $u \cojoin N(v)$ for some $v \in N_0 \cap X$, $v \neq x$: 

\medskip

Recall that $dist_G(v,y)=3$, say $(v,y',x',y)$ induce a $P_4$ in $G[N_0 \cup N_1]$ with $x',y' \in N_1$ (possibly $x'=x_1)$. 
Since $u$ contacts $N(x)$, $uy_1 \in E$ or $uw \in E$, $w \neq y_1$, $w \in N(x)$.

First assume that $uy_1 \in E$. If $x_1y' \in E$ then $(u,y_1,x_1,y',v)$ induce a $P_5$ with endpoint $u$. If $x_1y' \notin E$ and $x'y_1 \in E$ then 
$(u,y_1,x',y',v)$ induce a $P_5$ with endpoint $u$. However, if $x_1y' \notin E$ and $x'y_1 \notin E$ then $(u,y_1,x_1,y,x',y',v)$ induce a $P_7$ with endpoint $u$, which is impossible by Claim \ref{uinN2noendpointP6inN0N1}.

Next assume that $uy_1 \notin E$, say $uw \in E$, $w \neq y_1$, $w \in N(x)$. Since by Claim \ref{uinN2noendpointP6inN0N1}, $(u,w,x,y_1,x_1,y)$ does not induce a $P_6$ in $G$, we have $wx_1 \in E$. If $x_1y' \in E$ then $(u,w,x_1,y',v)$ induce a $P_5$ with endpoint $u$, and if $x_1y' \notin E$ but $wx' \in E$ then  
$(u,w,x',y',v)$ induce a $P_5$ with endpoint $u$. Finally, assume that $wx' \notin E$ and $x_1y' \notin E$. But then again $(u,w,x_1,y,x',y',v)$ induce a $P_7$ with endpoint $u$, which is impossible by Claim \ref{uinN2noendpointP6inN0N1}.

\medskip

Thus, for every $u \in N_2 \cap X$, $u$ is the endpoint of a $P_5$ whose remaining vertices are in $N_0 \cup N_1$.
Analogously, $(i)$ can be done if $u \in N_2 \cap Y$. 

\medskip

\noindent
$(ii)$: Without loss of generality, assume that $u \in N_2 \cap X$. Since $u$ is black, $u \cojoin N(y)$ since $N(y) \subset X$, but $u$ contacts $N(x)$ since $u$
contacts all $N(v)$ with $v \in N_0 \cap X$. 

If $uy_1 \in E$ then $(u,y_1,x_1,y)$ induce a $P_4$ with endpoint $u$. 
If $uy_1 \notin E$ and $uy' \in E$ for $y' \in N(x)$ and $y'x_1 \notin E$ then $(u,y',x,y_1,x_1,y)$ induce a $P_6$ with endpoint $u$, which is impossible by Claim \ref{uinN2noendpointP6inN0N1}. Thus, $y'x_1 \in E$ and then $(u,y',x,y_1)$ induce a $P_4$ with endpoint $u$.

\medskip

Thus, for every $u \in N_2 \cap X$, $u$ is the endpoint of a $P_4$ whose remaining vertices are in $N_0 \cup N_1$.
Analogously, $(ii)$ can be done if $u \in N_2 \cap Y$. 

\medskip

\noindent
$(iii)$: By $(ii)$ and since $G$ is $P_8$-free, $u$ is not the endpoint of a $P_5$ $(u,v_3,v_4,v_5,v_6)$ with $uv_3 \in E$, $v_3v_4 \in E$, $v_4v_5 \in E$, $v_5v_6 \in E$, and $v_3 \in N_3$, $v_4 \in N_3 \cup N_4$, 
$v_5 \in N_3 \cup N_4 \cup N_5$, $v_6 \in N_3 \cup N_4 \cup N_5 \cup N_6$. In particular, $u$ is not the endpoint of a $P_5$ $(u,v_3,v_4,v_5,v_6)$ with $v_i \in N_i$, $3 \le i \le 6$. Thus, $N_6=\emptyset$.
  
Analogously, $u$ is not the endpoint of a $P_5$ $(u,v_3,v_4,v_5,v_6)$ with $v_3 \in N_3$, $v_4 \in N_4$, and $v_5,v_6 \in N_5$.  
Thus, $N_5$ is independent. 

Finally, $u$ is not the endpoint of a $P_5$ $(u,v_3,v_4,v_5,v_6)$ with $v_4,v_5 \in N_4$ and $v_6 \in N_5$. Thus, every edge in $N_4$ does not contact $N_5$.

\medskip

Then Lemma \ref{N2endpointP4} is shown.
\qed

\medskip

In the next section, we assume that every $u \in N_2$ is the endpoint of a $P_5$ whose remaining vertices are in $N_0 \cup N_1$. 

\section{When every $u \in N_2$ is the endpoint of a $P_5$ whose remaining vertices are in $N_0 \cup N_1$}\label{N2endpointP5N0N1}

\subsection{General remarks}

\begin{lemma}\label{N2endpointP5}
If every $u \in N_2$ is the endpoint of a $P_5$ whose remaining vertices are in $N_0 \cup N_1$ then we have: 
\begin{itemize}
\item[$(i)$] Every vertex in $N_2$ is no endpoint of any $P_4$ whose remaining vertices are in $N_i$, $i \ge 3$. 
\item[$(ii)$] $N_5=\emptyset$, $N_4$ is independent, and every edge in $N_3$ does not contact $N_4$.
\end{itemize}
\end{lemma}

\noindent
{\bf Proof.} 
$(i)$: Suppose to the contrary that $u \in N_2$ is the endpoint of a $P_4$ $(u,u_3,u_4,u_5)$ with $u_3 \in N_3$, $u_4 \in N_3 \cup N_4$, $u_5 \in N_3 \cup N_4 \cup N_5$. Recall that $u$ is the endpoint of a $P_5$ $P$ whose remaining vertices are in $N_0 \cup N_1$. But then $u_5,u_4,u_3$ and the $P_5$ $P$ with endpoint $u$ induce a $P_8$ in $G$, which is a contradiction. Thus, $u$ is no endpoint of any $P_4$ whose remaining vertices are in $N_i$, $i \ge 3$.

\medskip

\noindent
$(ii)$: By $(i)$, every $u \in N_2$ is no endpoint of any $P_4$ whose remaining vertices are in $N_i$, $i \ge 3$.

\medskip

If $N_5 \neq \emptyset$, say $u_i \in N_i$, $2 \le i \le 5$, with $u_2u_3 \in E$, $u_3u_4 \in E$ and $u_4u_5 \in E$, 
then $u_2$ is the endpoint of such a $P_4$ $(u_2,u_3,u_4,u_5)$, which is impossible by $(i)$. Thus, $N_5=\emptyset$.

\medskip

Analogously, if $N_4$ is not independent, say $u_4v_4 \in E$ with $u_4,v_4 \in N_4$, then let $u_3 \in N_3$ be a neighbor of $u_4$. Clearly, $u_3v_4 \notin E$ since $G$ is bipartite. Recall that there is a neighbor $u_2 \in N_2$ of $u_3$. But then $u_2$ is the endpoint of such a $P_4$ $(u_2,u_3,u_4,v_4)$, which is impossible by $(i)$. Thus, $N_4$ is independent. 

\medskip

Finally, if there is an edge $u_3v_3 \in E$ with $u_3,v_3 \in N_3$ which contacts $N_4$, say without loss of generality, $v_3v_4 \in E$ with $v_4 \in N_4$, then let $u_2u_3 \in E$ for $u_2 \in N_2$. But then $u_2$ is the endpoint of such a $P_4$ $(u_2,u_3,v_3,v_4)$, which is impossible by $(i)$.  
Thus, every edge in $N_3$ does not contact $N_4$. 

\medskip

Now, Lemma \ref{N2endpointP5} is shown.
\qed 

\begin{coro}\label{vinN3eitherN(v)inN3orinN4}
For every $v \in N_3$, either $N(v) \cap N_3=\emptyset$ or $N(v) \cap N_4=\emptyset$.
\end{coro}

\noindent
{\bf Proof.}
Recall that by (\ref{vinN3neighbinN3N4}), for every $v \in N_3$, $N(v) \cap (N_3 \cup N_4) \neq \emptyset$ (else $v$ is $D_{basis}$-forced), i.e., $v$ has a neighbor in $N_3 \cup N_4$.
If $|N(v) \cap (N_3 \cup N_4)|=1$ then either $N(v) \cap N_3=\emptyset$ or $N(v) \cap N_4=\emptyset$.
Now assume that $|N(v) \cap (N_3 \cup N_4)| \ge 2$, say $w_1,w_2 \in N(v) \cap (N_3 \cup N_4)$. 

\medskip

Suppose to the contrary that $w_1 \in N_3$ and $w_2 \in N_4$. Then $(v,w_1)$ induce a $P_2$ in $N_3$, and by Lemma \ref{N2endpointP5} $(ii)$, 
$(v,w_1)$ does not contact $N_4$, which is a contradiction.
Thus, either $N(v) \cap N_3=\emptyset$ or $N(v) \cap N_4=\emptyset$, and Corollary \ref{vinN3eitherN(v)inN3orinN4} is shown. 
\qed 

\medskip

Assume that $w_1,w_2 \in N_4$ are leaves in $G$ and $w_1,w_2$ have a common neighbor $v \in N_3$ (recall that $N_4$ is independent). 
If $w_1 \in D$ then by the e.d.s.\ property, $w_2 \notin D$ and $w_2$ does not have any $D$-neighbor in $G$, and analogously, if $w_2 \in D$ then $w_1$ does not have any $D$-neighbor in $G$. 
Thus, $w_1,w_2 \notin D$ and $v$ is $D_{basis}$-forced. Then $D_{basis}$ can be updated, i.e., $D_{basis}:=D_{basis} \cup \{v\}$. 
It can also lead to a contradiction, i.e., if for a vertex $u \in N_2$, $w_1,w_2,w_3,w_4 \in N_4 \cap N^2(u)$ are leaves and $v_1,v_2 \in N_3 \cap N(u)$ with 
$v_1w_i \in E$, $i \in \{1,2\}$, and $v_2w_j \in E$, $j \in \{3,4\}$, then $v_1$ and $v_2$ are $D_{basis}$-forced, which is a contradiction by the e.d.s.\ property, since $uv_1 \in E$ and $uv_2 \in E$. Now assume:
\begin{equation}\label{leavesnocommonneighb}
\mbox{ Leaves in } N_4 \mbox{ do not have any common neighbor in } N_3.
\end{equation}

\begin{clai}\label{u123inN2noP7}
There is no $P_7$ $(v_1,u_1,v_2,u_2,v_3,u_3,v_4)$ in $G$ with $u_i \in N_2$, $1 \le i \le 3$, and $v_i \in N_3$, $1 \le i \le 4$. 
\end{clai}

\noindent
{\bf Proof.} Suppose to the contrary that $P=(v_1,u_1,v_2,u_2,v_3,u_3,v_4)$ induce a $P_7$ in $G$ with $u_i \in N_2$, $1 \le i \le 3$, and 
$v_i \in N_3$, $1 \le i \le 4$. 
Recall that by (\ref{vinN3neighbinN3N4}), $v_1$ has a neighbor in $N_3 \cup N_4$ (else $v_1$ is $D_{basis}$-forced and $D_{basis}:=D_{basis} \cup \{v_1\}$); 
let $w \in N_3 \cup N_4$ with $v_1w \in E$. 

Since by Lemma \ref{N2endpointP5} $(i)$, $u_1$ is no endpoint of a $P_4$ $(u_1,v_1,w,v_3)$ and no endpoint of a $P_4$ $(u_1,v_1,w,v_4)$, we have $v_3w \notin E$ and 
$v_4w \notin E$.
Moreover, since $u_2$ is no endpoint of a $P_4$ $(u_2,v_2,w,v_1)$, we have $v_2w \notin E$.
But then $(w,v_1,u_1,v_2,u_2,v_3,u_3,v_4)$ induce a $P_8$ in $G$, which is a contradiction. Thus, Claim \ref{u123inN2noP7} is shown.
\qed

\medskip

By Lemma \ref{N2endpointP5}, we have:

\begin{coro}\label{uinN2noendpointP4N2N3N4}
If $u \in N_2$ and $w \in (N_3 \cup N_4) \cap N^2(u)$ then $N(w) \cap N_3 \subseteq N(u) \cap N_3$.  
\end{coro}

\noindent
{\bf Proof.} First assume that $w \in N_4 \cap N^2(u)$. 
Recall that by Lemma \ref{N2endpointP5} $(ii)$, $N_5=\emptyset$ and $N_4$ is independent. Clearly, since $w \in N_4 \cap N^2(u)$, 
$(u,v_1,w)$ induce a $P_3$ in $G$. Moreover, by Lemma \ref{N2endpointP5} $(i)$, there is no such $P_4$ $(u,v_1,w,v_2)$ with endpoint $u$ and $v_1,v_2 \in N_3$, 
i.e., $N(w) \cap N_3 \subseteq N(u) \cap N_3$. 

Next assume that $w \in N_3 \cap N^2(u)$ with $P_3$ $(u,v_1,w)$. Recall that by Lemma \ref{N2endpointP5} $(ii)$, every edge in $N_3$ does not contact $N_4$. 
Moreover, by Lemma \ref{N2endpointP5} $(i)$, there is no such $P_4$ $(u,v_1,w,v_2)$ with endpoint $u$ and $v_1,w,v_2 \in N_3$, 
i.e., $N(w) \cap N_3 \subseteq N(u) \cap N_3$. 

Thus, Corollary \ref{uinN2noendpointP4N2N3N4} is shown. 
\qed

\begin{coro}\label{uinN2winN3N4N(u)=N(w)}
If for $u \in N_2$ and $w \in (N_3 \cup N_4) \cap N^2(u)$ with $N(w) \cap N_3 = N(u) \cap N_3$ then $w$ is $D_{basis}$-excluded.  
\end{coro}

\noindent
{\bf Proof.}
Suppose to the contrary that for $w \in (N_3 \cup N_4) \cap N^2(u)$ with $N(w) \cap N_3 = N(u) \cap N_3$, $w \in D$. Then by the e.d.s.\ property, $N(u) \cap N_3 \cap D = \emptyset$, i.e., $u$ does not have any $D$-neighbor in $N_3$, and there is no such e.d.s.\ in $G$. Thus, $w$ is $D_{basis}$-excluded, and 
Corollary \ref{uinN2winN3N4N(u)=N(w)} is shown. 
\qed 

\medskip

Thus, by the e.d.s.\ property, $u$ must have a $D$-neighbor in $N(u) \cap N_3$, and $w$ has the same $D$-neighbor. 

\begin{coro}\label{nocommonpointN3N4}
For every $v \in N_3$ and $w \in N_4$, $v$ and $w$ do not have any common neighbor, i.e., $N(v) \cap N(w)=\emptyset$.  
\end{coro}

\noindent
{\bf Proof.}
First assume that $v \in N_3 \cap X$ and $w \in N_4 \cap Y$. Then clearly, since $G$ is bipartite, $v$ and $w$ do not have any common neighbor, and analogously, 
for $v \in N_3 \cap Y$ and $w \in N_4 \cap X$, $v$ and $w$ do not have any common neighbor.

Next assume that $v \in N_3 \cap X$ and $w \in N_4 \cap X$ or $v \in N_3 \cap Y$ and $w \in N_4 \cap Y$, say without loss of generality, 
$v \in N_3 \cap X$ and $w \in N_4 \cap X$. 
Since by Lemma~\ref{N2endpointP5} $(ii)$, $N_4$ is independent and every edge in $N_3$ does not contact $N_4$,  
we have that $v$ and $w$ do not have any common neighbor in $N_3 \cup N_4$, and Corollary \ref{nocommonpointN3N4} is shown. 
\qed

\begin{clai}\label{N(v1)subsetN(v2)v2excluded}
If $v_1,v_2 \in N_3 \cup N_4$ and $N(v_1) \cap (N_3 \cup N_4) \subseteq N(v_2) \cap (N_3 \cup N_4)$ then $v_2$ is $D_{basis}$-excluded.  
\end{clai}

\noindent
{\bf Proof.} By Corollary \ref{nocommonpointN3N4}, we have $v_1,v_2 \in N_3$ or $v_1,v_2 \in N_4$. Clearly, $v_1$ and $v_2$ have the same color (either black or white) since $N(v_1) \cap (N_3 \cup N_4) \subseteq N(v_2) \cap (N_3 \cup N_4)$.

First assume that $v_1,v_2 \in N_3$ and $N(v_1) \cap N_4 \subseteq N(v_2) \cap N_4$.
Suppose to the contrary that $v_2 \in D$. Then by the e.d.s.\ property, $N(v_2) \cap D=\emptyset$ and $v_1 \notin D$, but then, $v_1$ does not have any $D$-neighbor in $G$ (recall that $D \cap N_2 = \emptyset$), i.e., $G$ has no such e.d.s. Thus, $v_2$ is $D_{basis}$-excluded. 

Next assume that $v_1,v_2 \in N_4$ and $N(v_1) \cap N_3 \subseteq N(v_2) \cap N_3$. 
Suppose to the contrary that $v_2 \in D$. Then by the e.d.s.\ property, $N(v_2) \cap D=\emptyset$ and $v_1 \notin D$, but then, $v_1$ does not have any $D$-neighbor in $G$ (recall that by Lemma \ref{N2endpointP5} $(ii)$, $N_5 = \emptyset$ and $N_4$ is independent), 
i.e., $G$ has no such e.d.s. Then again, $v_2$ is $D_{basis}$-excluded, and Claim \ref{N(v1)subsetN(v2)v2excluded} is shown. 
\qed 

\begin{coro}\label{N(v1)=N(v2)}
The following statements hold:
\begin{itemize}
\item[$(i)$] If $v_1,v_2 \in N_3$ and $N(v_1) \cap (N_3 \cup N_4) = N(v_2) \cap (N_3 \cup N_4) =\{w\}$ then $w$ is $D_{basis}$-forced.   
\item[$(ii)$] If $v_1,v_2 \in N_3$ and $N(v_1) \cap (N_3 \cup N_4) = N(v_2) \cap (N_3 \cup N_4) =\{w_1,\ldots,w_{\ell}\}$, $\ell \ge 2$, and there are two such $N(w_i) \cap (N_3 \cup N_4) = N(w_j) \cap (N_3 \cup N_4) =\{v_1,v_2\}$, $1 \le i,j \le \ell$, $i \neq j$, then there is no such e.d.s.\ in $G$ 
\end{itemize}
\end{coro}

\noindent
{\bf Proof.}
$(i)$: Recall that by Claim \ref{N(v1)subsetN(v2)v2excluded}, $v_1,v_2$ are $D_{basis}$-excluded and must have a $D$-neighbor in $N_3 \cup N_4$. Thus, 
$N(v_1) \cap (N_3 \cup N_4) = N(v_2) \cap (N_3 \cup N_4) =\{w\}$, i.e., $w$ is $D_{basis}$-forced.

\medskip

\noindent
$(ii)$: Assume that $v_1,v_2 \in N_3$ and $N(v_1) \cap (N_3 \cup N_4) = N(v_2) \cap (N_3 \cup N_4) =\{w_1,\ldots,w_{\ell}\}$, $\ell \ge 2$. 
Recall again that by Claim \ref{N(v1)subsetN(v2)v2excluded}, $v_1,v_2$ are $D_{basis}$-excluded and must have a $D$-neighbor.

Without loss of generality, assume that $N(w_1) \cap N_3 = N(w_2) \cap N_3 =\{v_1,v_2\}$. 
Then either $w_1$ or $w_2$ is $D_{basis}$-excluded and does not have any $D$-neighbor, i.e., there is no such e.d.s.\ in $G$. 

\medskip

Thus, Corollary \ref{N(v1)=N(v2)} is shown.
\qed

\medskip

Then by Corollary \ref{N(v1)=N(v2)} $(i)$, $D_{basis}$ can be updated, i.e., $D_{basis}:=D_{basis} \cup \{w\}$.

\begin{coro}\label{N(v1)subN(v2)subN(v3)}
The following statements hold:
\begin{itemize}
\item[$(i)$] If $v_1,v_2 \in N_3$ with $N(v_1) \cap (N_3 \cup N_4) = \{w_1\}$ and $N(v_2) \cap (N_3 \cup N_4) =\{w_1,w_2\}$ such that  
$N(w_2) \cap N_3 = \{v_2\}$, then $v_1$ and $w_2$ are $D_{basis}$-forced.     
\item[$(ii)$] If $v_1,v_2 \in N_3$ and $N(v_1) \cap (N_3 \cup N_4) = \{w_1\}$, 
$N(v_2) \cap (N_3 \cup N_4) =\{w_1,w_2,w_3\}$, with $N(w_3) \cap N_3 =\{v_2\}$, then there is no such e.d.s.\ in $G$. 
\end{itemize}
\end{coro}

\noindent
{\bf Proof.}
$(i)$: Recall that by Claim \ref{N(v1)subsetN(v2)v2excluded}, $v_2$ is $D_{basis}$-excluded and $v_2$ must have a $D$-neighbor in $G$. 
Moreover, recall that $N(w_2) \cap N_3 = \{v_2\}$. If $w_1 \in D$ and $v_2 \notin D$ then $w_2$ does not have any $D$-neighbor in $G$.    
Thus, $w_1$ is $D_{basis}$-excluded and $w_2$ is $D_{basis}$-forced. Since $N(v_1) \cap (N_3 \cup N_4) = \{w_1\}$ and $w_1 \notin D$, $v_1$ is $D_{basis}$-forced.

\medskip

\noindent
$(ii)$: Recall that by Claim \ref{N(v1)subsetN(v2)v2excluded}, $v_2$ is $D_{basis}$-excluded and must have a $D$-neighbor in $G$. 
Moreover, since $N(w_3) \cap N_3 =\{v_2\}$ and $v_2 \notin D$, $w_3$ is $D_{basis}$-forced, and by the e.d.s.\ property, $w_1,w_2 \notin D$. 
Since $N(v_1) \cap (N_3 \cup N_4) = \{w_1\}$, $v_1$ is $D_{basis}$-forced, and since $v_2 \notin D$, $w_2$ must have a $D$-neighbor $v_3 \in D \cap N_3$.
Let $u \in N_2$ with $uv_1 \in E$. Since by Lemma~\ref{N2endpointP5} $(i)$, $u$ is no endpoint of a $P_4$ $(u,v_1,w_1,v_2)$, we have $uv_2 \in E$. 
Recall that $v_1 \in D$, and by the e.d.s.\ property, $u$ has only one $D$-neighbor in $G$, i.e., $uv_3 \notin E$.  
But then $(u,v_2,w_2,v_3)$ induce a $P_4$ with endpoint $u$, which is a contradiction, and there is no such e.d.s.\ in $G$. 

\medskip

Thus, Corollary \ref{N(v1)subN(v2)subN(v3)} is shown.
\qed

\medskip

Then by Corollary \ref{N(v1)subN(v2)subN(v3)} $(i)$, $D_{basis}$ can be updated, i.e., $D_{basis}:=D_{basis} \cup \{v_1,w_2\}$.







\begin{clai}\label{v1inDN3v2Dneighbforced}
If for $u \in N_2$, $N(u) \cap N_3 = \{v_1,\ldots,v_r\}$, $r \ge 2$, and $v_1 \in D$ then for every $i \ge 2$,
$|(N(v_i) \setminus N(v_1)) \cap (N_3 \cup N_4)|=1$, say $(N(v_i) \setminus N(v_1)) \cap (N_3 \cup N_4)=\{w_i\}$, and vertex $w_i$ is 
$D_{basis} \cup \{v_1\}$-forced.  
\end{clai}

\noindent
{\bf Proof.} Without loss of generality, assume that $i=2$. Since $v_1 \in D$, we have $v_2 \notin D$, and $v_2$ must have a $D$-neighbor in 
$(N_3 \cup N_4) \setminus N(v_1)$. If $(N(v_2) \setminus N(v_1)) \cap (N_3 \cup N_4) = \emptyset$ then by the e.d.s.\ property, $v_2$ does not have any $D$-neighbor, and there is no such e.d.s.\ in $G$. Thus, $(N(v_2) \setminus N(v_1)) \cap (N_3 \cup N_4) \neq \emptyset$. 

\medskip

Suppose to the contrary that $|(N(v_2) \setminus N(v_1)) \cap (N_3 \cup N_4)| \ge 2$, say $w_2,w'_2 \in (N(v_2) \setminus N(v_1)) \cap (N_3 \cup N_4)$, and 
without loss of generality, let $w_2 \in D$. Then $w'_2 \notin D$ and $w'_2$ must have a $D$-neighbor. Recall that by Lemma~\ref{N2endpointP5} $(ii)$, 
$N_5=\emptyset$, $N_4$ is independent, and every edge in $N_3$ does not contact $N_4$. Then either $w_2,w'_2 \in N_3$ or $w_2,w'_2 \in N_4$. 
   
First assume that $w_2,w'_2 \in N_4$. Then $w'_2$ must have a $D$-neighbor in $N_3$, say $w'_2v_3 \in E$ with $v_3 \in D \cap N_3$. But then by the e.d.s.\ property, $uv_3 \notin E$, i.e., $(u,v_2,w'_2,v_3)$ induce a $P_4$ with endpoint $u$, which is impossible by Lemma~\ref{N2endpointP5} $(i)$.

Next assume that $w_2,w'_2 \in N_3$. Then again $w'_2$ must have a $D$-neighbor in $N_3$, say $w'_2v_3 \in E$ with $v_3 \in D \cap N_3$. 
But then again $uv_3 \notin E$, i.e., $(u,v_2,w'_2,v_3)$ induce a $P_4$ with endpoint $u$, which is impossible by Lemma~\ref{N2endpointP5} $(i)$.

Thus, $|(N(v_2) \setminus N(v_1)) \cap (N_3 \cup N_4)|=1$, say $(N(v_2) \setminus N(v_1)) \cap (N_3 \cup N_4)=\{w_2\}$, i.e., $w_2$ is $D_{basis} \cup \{v_1\}$-forced, and Claim \ref{v1inDN3v2Dneighbforced} is shown. 
\qed
 
\begin{coro}\label{N(v2)minusN(v1)ge2v1excluded}
For $u \in N_2$ and $v_i,v_j \in N(u) \cap N_3$, $i \neq j$, with $|(N(v_j) \setminus N(v_i)) \cap (N_3 \cup N_4)| \ge 2$, we have: 
\begin{itemize}
\item[$(i)$] $v_i$ is $D_{basis}$-excluded.     
\item[$(ii)$] If $N(v_i) \cap (N_3 \cup N_4)=\{w_i\}$ then $w_i$ is $D_{basis}$-forced, or it leads to a contradiction. 
\end{itemize}
\end{coro}

\noindent
{\bf Proof.} 
Assume that $v_1,v_2 \in N(u) \cap N_3$ with $|(N(v_2) \setminus N(v_1)) \cap (N_3 \cup N_4)| \ge 2$, say 
$w_2,w'_2 \in (N(v_2) \setminus N(v_1)) \cap (N_3 \cup N_4)$. 

\medskip

\noindent
$(i)$: Recall that by Claim \ref{v1inDN3v2Dneighbforced}, we have: If $v_1 \in D$ then $|(N(v_2) \setminus N(v_1)) \cap (N_3 \cup N_4)|=1$. 
Since $|(N(v_j) \setminus N(v_i)) \cap (N_3 \cup N_4)| \ge 2$, $v_i$ is $D_{basis}$-excluded.    

\medskip

\noindent
$(ii)$: Assume that $N(v_1) \cap (N_3 \cup N_4)=\{w_1\}$. Then by $(i)$, $v_1$ is $D_{basis}$-excluded and $v_1$ must have a $D$-neighbor. 
Since $D \cap N_2=\emptyset$, we have that $w_1$ is $D_{basis}$-forced (else there is no such e.d.s.\ in $G$ with $D_{basis}$). 
Moreover, if $N(v_1) \cap (N_3 \cup N_4)=\{w_1\}$ as well as $N(v'_1) \cap (N_3 \cup N_4)=\{w'_1\}$ and $w_1,w'_1$ have a common neighbor in $N_2 \cup N_3$ then there is no such e.d.s. with $D_{basis}$. 

\medskip

Thus, Corollary \ref{N(v2)minusN(v1)ge2v1excluded} is shown.   
\qed 

\medskip

If by Corollary \ref{N(v2)minusN(v1)ge2v1excluded} $(ii)$, $w_1$ is $D_{basis}$-forced then $D_{basis}$ can be updated, i.e., $D_{basis}:=D_{basis} \cup \{w_1\}$.
Now assume that for every such $v_i \in N(u) \cap N_3$, $|N(v_i) \cap (N_3 \cup N_4)| \ge 2$. 

\begin{coro}\label{N(v1)N(v2)ge2nocommonneighb}
For $v_1,v_2 \in N(u) \cap N_3$ with $|N(v_i) \cap (N_3 \cup N_4)| \ge 2$, $i \in \{1,2\}$, and $N(v_1) \cap N(v_2) \cap (N_3 \cup N_4)=\emptyset$, we have:
\begin{itemize}
\item[$(i)$] $v_1$ and $v_2$ must have distinct $D$-neighbors, say $D \cap N(v_i) \cap (N_3 \cup N_4)=\{w_i\}$, $i \in \{1,2\}$. 
\item[$(ii)$] There is a common $D$-neighbor $v \in D \cap N_3$ with $v \join (N(v_i) \setminus \{w_i\}) \cap (N_3 \cup N_4))$, $i \in \{1,2\}$. 
\item[$(iii)$] $N(v_i) \cap N_3=\emptyset$, $i \in \{1,2\}$.   
\end{itemize}
\end{coro}

\noindent
{\bf Proof.} $(i)$: Assume that for $v_1,v_2 \in N(u) \cap N_3$, $|(N(v_i) \cap (N_3 \cup N_4)| \ge 2$, $i \in \{1,2\}$, and there is no common neighbor between $v_1$ and $v_2$ in $N_3 \cup N_4$. 

By Corollary \ref{N(v2)minusN(v1)ge2v1excluded} $(i)$, $v_1,v_2$ are $D_{basis}$-excluded, and since there is no common neighbor between $v_1$ and $v_2$ in 
$N_3 \cup N_4$, $v_1$ and $v_2$ must have distinct $D$-neighbors in $N_3 \cup N_4$, say without loss of generality, 
$w_1 \in D \cap N(v_1)$ and $w_2 \in D \cap N(v_2)$.

\medskip

\noindent
$(ii)$: By $(i)$, we have $w_i \in D \cap N(v_i)$, $i \in \{1,2\}$, with $w_i \in N_3 \cup N_4$. Then by the e.d.s.\ property, every vertex in 
$N(v_i) \cap (N_3 \cup N_4) \setminus \{w_i\}$, $i \in \{1,2\}$, must have a $D$-neighbor. First assume that $v \in D \cap N_3$ with $vw'_1 \in E$, 
$w'_1 \in N(v_1) \cap (N_3 \cup N_4) \setminus \{w_1\}$. 
Recall that by Lemma~\ref{N2endpointP5} $(i)$, $u$ is no endpoint of such a $P_4$ $(u,v_1,w'_1,v)$, i.e., $uv \in E$. By the e.d.s.\ property, $u$ has exactly one 
$D$-neighbor $v \in D \cap N_3$. Then there is no second $D$-neighbor for $N(v_i) \cap (N_3 \cup N_4) \setminus \{w_i\}$, $i \in \{1,2\}$, i.e., there is a common 
$D$-neighbor $v \in D \cap N_3$ with $v \join (N(v_i) \setminus \{w_i\}) \cap (N_3 \cup N_4)$, $i \in \{1,2\}$. 

\medskip

\noindent
$(iii)$:  
First suppose to the contrary that $N(v_i) \cap N_3 \neq \emptyset$, $i \in \{1,2\}$. Let $(w_i,v_i,w'_i)$, $i \in \{1,2\}$, induce a $P_3$ in $N_3$, and 
$w_i \in D \cap N(v_i)$, $i \in \{1,2\}$. Assume that $u'w_1 \in E$ with $u' \in N_2$. Then by the e.d.s.\ property, $u'w_2 \notin E$. 
Recall that by $(ii)$, $v \in D \cap N_3$ is a common $D$-neighbor, i.e., $vw'_1 \in E$ and $vw'_2 \in E$. Since by Lemma~\ref{N2endpointP5} $(i)$, $(u,v_1,w'_1,v)$ does not induce a $P_4$ with endpoint $u$, we have $uv \in E$. 
Moreover, since $(u',w_1,v_1,w'_1)$ does not induce a $P_4$ with endpoint $u$, we have $u'w'_1 \in E$, and since $(u',w'_1,v,w'_2)$ does not induce a $P_4$ with endpoint $u$, we have $u'w'_2 \in E$. But then by the e.d.s.\ property, $u'w_2 \notin E$, and $(u',w'_2,v_2,w_2)$ induce a $P_4$ with endpoint $u$, which is a contradiction. 

\medskip

Next suppose to the contrary that $N(v_1) \cap N_3 \neq \emptyset$ and $N(v_2) \cap N_3 = \emptyset$, say $(w_1,v_1,w'_1)$ induce a $P_3$ in $N_3$, and $(w_2,v_2,w'_2)$ induce a $P_3$ with $v_2 \in N_3$ and $w_2,w'_2 \in N_4$. Moreover, $w_1,w_2 \in D$. 
Recall that by $(ii)$, there is a common $D$-neighbor $v \in D \cap N_3$ with $v \join (N(v_i) \setminus \{w_i\}) \cap (N_3 \cup N_4))$, $i \in \{1,2\}$. 
However by Lemma~\ref{N2endpointP5} $(ii)$, every $P_2$ in $N_3$ does not contact $N_4$, i.e., since $vw'_1 \in E$ then $vw'_2 \notin E$, which is impossible.  
 
\medskip

Thus, Corollary \ref{N(v1)N(v2)ge2nocommonneighb} is shown. 
\qed

\medskip

If every $v_i \in N(u) \cap N_3$, $1 \le i \le r$, is $D_{basis}$-excluded then there is no such e.d.s.\ in $G$ with $D_{basis}$.
Moreover, it can also lead to the case that $G$ has no such e.d.s.\ in $G$: 
\begin{itemize}
\item[$(i)$] If $v_2$ contacts only $N(v_1) \cap (N_3 \cup N_4)$ and $v_1 \in D$ then $v_2$ has no $D$-neighbor, and there is no such e.d.s.\ in $G$. 
\item[$(ii)$] If $N_4 \cap N^2(u)=\{w\}$ with $w \join N(u) \cap N_3$ then there is no such e.d.s.\ in $G$. 
\item[$(iii)$] If two vertices in $(N^2(u) \setminus N(v_1)) \cap N_4$ have a common neighbor in $N_3$ then by Claim \ref{v1inDN3v2Dneighbforced}, there is no such 
e.d.s.\ in $G$. 
\end{itemize}

\subsection{Components $Q$ in $G[N_2 \cup N_3 \cup N_4]$}\label{sectQN2N3N4}

Let $N^X_i:=N_i \cap X$ and $N^Y_i:=N_i \cap Y$, $2 \le i \le 4$. 
For every component $Q$ in $G[N_2 \cup N_3 \cup N_4]$, the e.d.s.\ problem for $G$ can be done independently. 

\medskip

If there is an e.d.s.\ $D$ in $Q$ then for a component $Q'$ with $V(Q') \cap N^X_i = V(Q) \cap N^X_i$, $V(Q') \cap N^Y_i = V(Q) \cap N^Y_i$, $2 \le i \le 4$, 
and $(V(Q') \cap N^X_2) \join (V(Q') \cap N^Y_2)$, $Q'$ has the same e.d.s.\ $D$ in $Q'$ as in $Q$. Thus assume: 
\begin{equation}\label{(N2QXjoinY)}
(V(Q) \cap N^X_2) \join (V(Q) \cap N^Y_2).
\end{equation}

\subsubsection{Components $Q$ in $G[N_2 \cup N_3 \cup N_4]$ with independent $V(Q) \cap N_3$}\label{sectQN3indep}

Let $Q$ be a component in $G[N_2 \cup N_3 \cup N_4]$,  and assume that $V(Q) \cap N_3$ is independent, i.e., there is no edge in $V(Q) \cap N_3$, and for every vertex $v \in V(Q) \cap N_3$, $v$ does not have any neighbor in $N_3$. 
Now assume that every vertex in $V(Q) \cap N_3$ has a neighbor in $N_4$ (recall (\ref{vinN3neighbinN3N4})). 
Moreover, recall that by Lemma \ref{N2endpointP5} $(ii)$, $N_5=\emptyset$ and $N_4$ is independent.
 
\begin{coro}\label{2DinN3dist3or4}
For every $v_1,v_2 \in D \cap V(Q) \cap N_3$, we have:
\begin{itemize}
\item[$(i)$] If $v_1 \in X$ and $v_2 \in Y$ then $dist_Q(v_1,v_2)=3$, i.e., $(v_1,u_1,u_2,v_2)$ induce a $P_4$ with $u_1,u_2 \in N_2$.
\item[$(ii)$] If $v_1,v_2 \in X$ or $v_1,v_2 \in Y$ then $dist_Q(v_1,v_2)=4$, i.e., $(v_1,u_1,v,u_2,v_2)$ induce a $P_5$ with $u_1,u_2 \in N_2$ and 
$v \in N_2 \cup N_3$. 
\end{itemize}
\end{coro}

\noindent
{\bf Proof.}
Let $v_i  \in D \cap V(Q) \cap N_3$, $i \in \{1,2\}$. Since $v_i$ is not $D_{basis}$-forced and in this section, $N_3$ is independent, $v_i$ contacts $N_4$, say $v_iw_i \in E$ with $w_i \in N_4$, $i \in \{1,2\}$. 

\medskip

\noindent
$(i)$: Assume that $v_1 \in D \cap N^X_3$ and $v_2 \in D \cap N^Y_3$. Let $u_1v_1 \in E$ with $u_1 \in N^Y_2$ and $u_2v_2 \in E$ with 
$u_2 \in N^X_2$. By (\ref{(N2QXjoinY)}), $u_1u_2 \in E$, and thus, $dist_Q(v_1,v_2)=3$, i.e., $(v_1,u_1,u_2,v_2)$ induce a $P_4$ with $u_1,u_2 \in N_2$.  

\medskip 

\noindent
$(ii)$: Without loss of generality, assume that $v_1,v_2 \in D \cap N^X_3$. Let $u_iv_i \in E$ with $u_i \in N^Y_2$, $i \in \{1,2\}$.
By the e.d.s.\ property, $u_1 \neq u_2$, and if $V(Q) \cap N^X_2 \neq \emptyset$ then by (\ref{(N2QXjoinY)}), $u_1,u_2$ have a common neighbor $v \in N^X_2$, i.e., $(v_1,u_1,v,u_2,v_2)$ induce a $P_5$ with $u_1,u_2 \in N_2$ and $v \in N_2$. 

Now assume that $V(Q) \cap N^X_2 = \emptyset$. 
Since by Claim \ref{u123inN2noP7}, there is no such $P_7$ $(v_1,u_1,v,u,v',u_2,v_2)$ in $V(Q) \cap (N_2 \cup N_3)$ with $v,v' \in N^X_3$, $u \in N^Y_2$, 
$u_1,u_2$ have a common neighbor $v \in N^X_3$, i.e., $(v_1,u_1,v,u_2,v_2)$ induce a $P_5$ with $u_1,u_2 \in N_2$ and $v \in N_3$. 
Then $dist_Q(v_1,v_2)=4$, and analogously, for $v_1,v_2 \in D \cap N^Y_3$, $dist_Q(v_1,v_2)=4$. 

\medskip 

Thus, Corollary \ref{2DinN3dist3or4} is shown.
\qed
 
\begin{lemma}\label{QN3indepedspol}
If for component $Q$ in $G[N_2 \cup N_3 \cup N_4]$, $V(Q) \cap N_3$ is independent then the e.d.s.\ problem for $Q$ can be done in polynomial time.   
\end{lemma}

\noindent
{\bf Proof.} Recall $N^X_i:=N_i \cap X$ and $N^Y_i:=N_i \cap Y$, $2 \le i \le 4$. 

\medskip 

\noindent
{\bf Case 1.} $V(Q) \cap N^X_2=\emptyset$ or $V(Q) \cap N^Y_2=\emptyset$: 

\medskip 

Without loss of generality, assume that $V(Q) \cap N^Y_2=\emptyset$ and $V(Q) \cap N^X_2 \neq \emptyset$. 
In particular, if there is a $u \in N^X_2$ with $u \join V(Q) \cap N^Y_3$ then by the e.d.s.\ property, there is exactly one $D$-vertex in $N^Y_3$, and 
the e.d.s.\ problem for $Q$ can be done in polynomial time. 
Now assume that there is no $u \in N^X_2$ with $u \join V(Q) \cap N^Y_3$.
If $|D \cap V(Q) \cap N_3| \le 2$ then the e.d.s.\ problem for $Q$ can be done in polynomial time. Thus assume that $|D \cap V(Q) \cap N_3| \ge 3$. 

\medskip

Let $u_1,u_2$ in $N^X_2$ with special neighbors $v_1,v_2 \in V(Q) \cap N^Y_3$ induce a $2P_2$, i.e., $u_1v_1 \in E$, $u_2v_2 \in E$, and $u_1v_2 \notin E$, $u_2v_1 \notin E$. 

Recall that $V(Q) \cap N_3$ is independent and $V(Q) \cap N^Y_2=\emptyset$, i.e., $v_1,v_2$ do not have any neighbors in $N_3$ and 
by (\ref{vinN3neighbinN3N4}), $v_1,v_2$ must have neighbors in $N^X_4$ (else $v_1$ or $v_2$ is $D_{basis}$-forced, and $D_{basis}$ can be updated). 
Thus assume that $v_1w_1 \in E$ and $v_2w_2 \in E$ for $w_1,w_2 \in N^X_4$. 

Recall that by Lemma \ref{N2endpointP5} $(i)$, $u_1$ is no endpoint of any $P_4$ whose remaining vertices are in $N_3 \cup N_4$, i.e., $v_1,v_2$ do not have any common neighbor $w \in N^X_4$ (else $(u_1,v_1,w,v_2)$ induce a $P_4$, which is a contradiction), i.e., $w_1 \neq w_2$, $v_1w_2 \notin E$, $v_2w_1 \notin E$, and 
$(u_1,v_1,w_1)$, $(u_2,v_2,w_2)$ induce a $2P_3$ in $Q$.  

\medskip

We first claim that there is no path in $Q[N^Y_3 \cup N^X_4]$ between $v_1$ and $v_2$: 

\medskip

Since $G$ is $P_8$-free bipartite, the distance between $v_1$ and $v_2$ in $Q[N_3 \cup N_4]$ is at most 6. Recall that $v_1,v_2$ do not have any common neighbor 
$w \in N_4$. 

\medskip

If there is a $P_5$-path $(v_1,w_1,v,w_2,v_2)$ with $w_1,w_2 \in N^X_4$ and $v \in N^Y_3$ then $u_1v \notin E$ (else $(u_1,v,w_2,v_2)$ induce a $P_4$, which is a contradiction), and analogously, $u_2v \notin E$. But then, $(u_1,v_1,w_1,v)$ induce a $P_4$, which is a contradiction. 

\medskip

Moreover, if there is a $P_7$-path $(v_1,w_1,v,w,v',w_2,v_2)$ with $w_1,w,w_2 \in N^X_4$ and $v,v' \in N^Y_3$ then $u_1v' \notin E$ (else $(u_1,v',w_2,v_2)$ induce a $P_4$ with endpoint $u_1$, which is a contradiction), and thus, $u_1v \notin E$ (else $(u_1,v,w,v')$ induce a $P_4$ with endpoint $u_1$, which is a contradiction), 
but then $(u_1,v_1,w_1,v)$ induce a $P_4$ with endpoint $u_1$, which is a contradiction.

\medskip

Thus, there is no such path in $Q[N^Y_3 \cup N^X_4]$ between $v_1$ and $v_2$. 

\medskip

Since $v_1,v_2 \in V(Q) \cap N^Y_3$, there is a path in $N^X_2 \cup N^Y_3$ between $v_1$ and $v_2$. Recall that by Claim \ref{u123inN2noP7}, there is no $P_7$ in 
$N_2 \cup N_3$ with endpoints $v_1,v_2$. Thus, since $(u_1,v_1)$, $(u_2,v_2)$, induce a $2P_2$ in $Q$,  
there is a $P_5$ $(v_1,u_1,v,u_2,v_2)$ in $V(Q) \cap (N^X_2 \cup N^Y_3)$ with $v \in N^Y_3$, i.e., 
$dist_Q(v_1,v_2)=4$ and $u_1,u_2$ have a common neighbor $v \in V(Q) \cap N^Y_3$. 

\medskip

Recall that $|D \cap V(Q) \cap N_3| \ge 3$, i.e., let $u_1,u_2,u_3 \in V(Q) \cap N^X_2$ with special neighbors $v_i \in V(Q) \cap  N^Y_3$ (possibly $v_i \in D$)
for $u_i$, $1 \le i \le 3$. 

Recall that $u_1,u_2$ have a special neighbor $v_{1,2} \in N^Y_3$, and $u_2,u_3$ have a special neighbor $v_{2,3} \in N^Y_3$. If $v_{1,2}u_3 \notin E$ and 
$v_{2,3}u_1 \notin E$ then $(v_1,u_1,v_{1,2},u_2,v_{2,3},u_3,v_3)$ induce a $P_7$ in $N_2 \cup N_3$, which is a contradiction by Claim \ref{u123inN2noP7}. 
Thus, there is a common neighbor of $u_1,u_2,u_3$ in $V(Q) \cap  N^Y_3$, and   
in general, all $u_i \in V(Q) \cap N^X_2$ have a common neighbor $v \in V(Q) \cap  N^Y_3$. Then either $v \in D$, and by Claim \ref{v1inDN3v2Dneighbforced}, 
the e.d.s.\ problem for $Q$ can be done in polynomial time, or $v \notin D$ and $v$ must have a $D$-neighbor in $N^X_4$, i.e., $vw \in E$ for 
$w \in D \cap V(Q) \cap N^X_4$, and every $u_i$ has a special $D$-neighbor which can also be done in polynomial time.   

\medskip

Analogously, the e.d.s.\ problem for $Q$ can be done in polynomial time when $V(Q) \cap N^Y_2 \neq \emptyset$ and $V(Q) \cap N^X_2 = \emptyset$.

\medskip

\noindent
{\bf Case 2.} $V(Q) \cap N^X_2 \neq \emptyset$ and $V(Q) \cap N^Y_2 \neq \emptyset$:

\medskip

By (\ref{(N2QXjoinY)}), $(V(Q) \cap N^X_2) \join (V(Q) \cap N^Y_2)$, i.e., for $v_1 \in D \cap N^X_3$ and $v_2 \in D \cap N^Y_3$, $dist_Q(v_1,v_2)=3$. 
Then the e.d.s.\ problem can be done independently for $V(Q) \cap N^X_2$ and for $V(Q) \cap N^Y_2$.    

\medskip

Thus, the e.d.s.\ problem can be done in polynomial time for $Q$ as in Case 1, and Lemma \ref{QN3indepedspol} is shown.  
\qed
 
\subsubsection{Components $Q$ in $G[N_2 \cup N_3 \cup N_4]$ with $N_3$-edges}\label{sectN3edges}

A component $K$ in $G[N_3]$ is {\em nontrivial} if $K$ has an edge (otherwise, $|V(K) \cap N_3|=1$ and $K$ is trivial in $G[N_3]$).

\begin{clai}\label{nontrivialcompinN3}
For every nontrivial component $K=(X_K,Y_K,E_K)$ in $N_3$, we have:
\begin{itemize}
\item[$(i)$] $N(K) \cap N_4 = \emptyset$. 
\item[$(ii)$] For every edge $(v,w)$ in $K$ with $v \in N^X_3$, $w \in N^Y_3$ and its neighbors $u_v,u_w$ in $N_2$, we have 
$u_v \join X_K$ and $u_w \join Y_K$.
\item[$(iii)$] If there is a $P_4$ in $K$ then $|D \cap X_K|=1$, say $D \cap X_K=\{v\}$, as well as $|D \cap Y_K|=1$, say $D \cap Y_K=\{w\}$, and 
$v \join (Y_K \setminus \{w\})$ as well as $w \join (X_K \setminus \{v\})$. 
In particular, if $K$ is $P_4$-free then $|D \cap V(K)|=1$, say $D \cap V(K)=\{v\}$ and $v \join (V(K) \setminus \{v\})$. 
\end{itemize}
\end{clai}

\noindent
{\bf Proof.} 
$(i)$: Recall that by Lemma \ref{N2endpointP5} $(ii)$, every edge in $N_3$ does not contact $N_4$. 
Thus, $N(K) \cap N_4 = \emptyset$. 

\medskip

\noindent
$(ii)$: If $K$ is an edge, i.e., $P_2$, in $N_3$, say $V(K)=\{v,w\}$, i.e., $X_K=\{v\}$ and $Y_K=\{w\}$ then for the only edge $vw \in E$ in $K$ and its neighbors $u_v,u_w$ in $N_2$, $u_vv \in E$, $u_ww \in E$, we have $u_v \join X_K$ and $u_w \join Y_K$. 

\medskip

Now assume that every $P_2$ in $K$ is part of a $P_3$, say $(v,w,v')$ in $K$. 
Recall $u_vv \in E$ and $u_ww \in E$ for $u_v,u_w \in N_2$. Since by Lemma \ref{N2endpointP5} $(i)$, $u_v$ is no endpoint of a $P_4$ $(u_v,v,w,v')$, we have
 $u_vv' \in E$, and in general, $u_v \join X_K$. Analogously, for a $P_3$ $(w,v,w')$, we have $u_ww' \in E$, and in general, $u_w \join Y_K$.

\medskip

\noindent
$(iii)$: Let $P=(v,w,v',w')$ with $v,v' \in X_K$, $w,w' \in Y_K$ be a $P_4$ in $K$. Recall that $D \cap N_2=\emptyset$ and by Claim \ref{nontrivialcompinN3} $(i)$, every edge in $K$ has no neighbor in $N_4$. Thus, if there is no $D$-vertex in $K$ then there is no such e.d.s. Moreover, if $D \cap X_K=\emptyset$ or 
$D \cap Y_K=\emptyset$ then there is no such e.d.s. In particular, if $D \cap X_K=\emptyset$ then $w$ or $w'$ does not have any $D$-neighbor (if $w \in D$ then 
$w' \notin D$ and vice versa). Analogously, if $D \cap Y_K=\emptyset$ then $v$ or $v'$ does not have any $D$-neighbor (if $v \in D$ then 
$v' \notin D$ and vice versa).  
Now assume that $|D \cap X_K| \ge 1$ and $|D \cap Y_K| \ge 1$.

\medskip

Suppose to the contrary that $D \cap V(K) \ge 3$, i.e., either $|D \cap X_K| \ge 2$ or $|D \cap Y_K| \ge 2$. 
Without loss of generality, say $|D \cap X_K| \ge 2$, i.e., $v_1,v_2 \in D \cap X_K$; let $uv_1 \in E$ for $u \in N_2$. 
Recall that by Claim \ref{nontrivialcompinN3} $(ii)$, $u \join X_K$. But then it leads to a contradiction by the e.d.s.\ property. 
Thus, if there is a $P_4$ $P=(v,w,v',w')$ in $K$ then $|D \cap X_K| = 1$ as well as $|D \cap Y_K| = 1$, say $D \cap X_K=\{v\}$ and $D \cap Y_K=\{w'\}$. 
Since $D \cap N_2 = \emptyset$ and $N(K) \cap N_4 = \emptyset$, we have $v \join (Y_K \setminus \{w'\})$ as well as $w' \join (X_K \setminus \{v\})$
(else there is no e.d.s.\ in $K$).

\medskip

In particular, if $K$ is $P_4$-free then $|D \cap V(K)| = 1$, say $D \cap V(K)=\{v\}$ and $v \join (V(K) \setminus \{v\})$.

\medskip

Thus, Claim \ref{nontrivialcompinN3} is shown.
\qed 

\begin{coro}\label{N3P7fr}
If there is a $P_7$ $P=(x_1,y_1,x_2,y_2,x_3,y_3,x_4)$ in $K$ then there are two $D$-vertices $v,w \in D \cap N_3$ with $v \join \{y_1,y_2,y_3\}$ and 
$w \join \{x_1,x_2,x_3,x_4\}$. 
Analogously, if there is a $P_7$ $P=(y_1,x_1,y_2,x_2,y_3,x_3,y_4)$ in $K$ then there are two $D$-vertices $v,w \in D \cap N_3$ with 
$v \join \{y_1,y_2,y_3,y_4\}$ and $w \join \{x_1,x_2,x_3\}$.
\end{coro}  

\noindent
{\bf Proof.} For the $P_7$ $P=(x_1,y_1,x_2,y_2,x_3,y_3,x_4)$ in $N_3$, recall that $D \cap N_2=\emptyset$ and by Lemma \ref{N2endpointP5} $(ii)$, 
$P$ does not contact $N_4$. 
Let $K=(X_K,Y_K,E_K)$ be the component in $G[N_3]$ containing $P$. By Claim \ref{nontrivialcompinN3} $(iii)$, we have $|D \cap X_K|=1$, say $D \cap X_K=\{v\}$, as well as $|D \cap Y_K|=1$, say $D \cap Y_K=\{w\}$, and $v \join Y_K$ as well as $w \join X_K$. Thus, $x_i \notin D$, $1 \le i \le 4$, and $y_j \notin D$, $1 \le j \le 3$ (else there is an e.d.s.\ contradiction), and $v \join \{y_1,y_2,y_3\}$ as well as $w \join \{x_1,x_2,x_3,x_4\}$.  
Analogously, the same holds for a $P_7$ $P=(y_1,x_1,y_2,x_2,y_3,x_3,y_4)$ in $G[N_3]$. Thus, Corollary \ref{N3P7fr} is shown.
\qed

\medskip

By Claim \ref{nontrivialcompinN3} $(iii)$, it can also lead to a contradiction. For example, if the nontrivial component $K$ in $N_3$ is a $C_4$, i.e., $K$ is $P_4$-free, then there is no such e.d.s.: Let $K=(x_1,y_1,x_2,y_2)$ be a $C_4$ in $N_3$. If $x_1 \in D$ then $x_2 \notin D$ and $x_2$ does not have any $D$-neighbor, which is a contradiction, and similarly for $x_2 \in D$, $y_1 \in D$ or $y_2 \in D$. 

\medskip

Moreover, if the nontrivial component $K$ in $N_3$ is exactly a $P_7$ then there is no such e.d.s.: Let $K=(x_1,y_1,x_2,y_2,x_3,y_3,x_4)$ be such a $P_7$. 
Then by Claim \ref{nontrivialcompinN3} $(iii)$, $|D \cap X_K|=1$, as well as $|D \cap Y_K|=1$ but then there is no such e.d.s.\ (recall Corollary \ref{N3P7fr}).

\medskip
 
Then $K$ can be reduced: 
\begin{itemize}
\item If there is a $P_4$ in $K$ and by Claim \ref{nontrivialcompinN3} $(iii)$, $D \cap X_K=\{v\}$ and $D \cap Y_K=\{w\}$ then reduce $K$ to a $P_4$ $(v,w',v',w)$. 
\item If there is no $P_4$ in $K$ and by Claim \ref{nontrivialcompinN3} $(iii)$, $D \cap V(K)=\{v\}$ then reduce $K$ to a $P_2$ $(v,w)$.  
\end{itemize}

For every $P_2$ $(v,w)$ in $N_3$ with no other neighbors of $v,w$ in $N_3$, either $v \in D$ or $w \in D$.

For every $P_4$ $(v_1,w_1,v_2,w_2)$ in $N_3$, $v_1 \in D$ and $w_2 \in D$, i.e., $v_1,w_2$ are $D_{basis}$-forced and $D_{basis}$ can be updated as 
$D_{basis}:=D_{basis} \cup \{v_1,w_2\}$. 
Moreover, we have:
\begin{itemize}
\item[$(i)$] If $uv_i \in E$, $u'w_i \in E$, $i \in \{1,2\}$, with $u,u' \in N_2$ and there is a $P_2$ $(v_3,w_3)$ in $N_3$ 
with $uv_3 \in E$, $u'w_3 \in E$ then $v_3,w_3$ are $D_{basis}$-excluded and there is no such e.d.s.\ in $G$. 
\item[$(ii)$] If $uv_i \in E$, $u'w_i \in E$, $i \in \{1,2\}$, with $u,u' \in N_2$ and there is a $2P_2$ $(v_3,w_3)$, $(v_4,w_4)$ in $N_3$ with $uv_3 \in E$, $uv_4 \in E$ and $u''w_3 \in E$, $u''w_4 \in E$ then $v_3,v_4 \notin D$, i.e., $w_3,w_4 \in D$, which is a contradiction by the e.d.s.\ property.
\end{itemize}

Now assume that there is no such $P_4$ in $N_3$.  

\begin{coro}\label{N3no3P2uu'}
The following statements hold:
\begin{itemize}
\item[$(i)$] For $u,u' \in N_2$, there is no $3P_2$ $(v_1,w_1)$, $(v_2,w_2)$, $(v_3,w_3)$ in $(N(u) \cup N(u')) \cap N_3$.
\item[$(ii)$] If for $u,u' \in N_2$, there is a $2P_2$ $(v_1,w_1)$, $(v_2,w_2)$ in $N_3$ with $uv_i \in E$, $u'w_i \in E$, $i \in \{1,2\}$, and $u$ contacts a third $P_2$ $(v_3,w_3)$ in $N_3$ then $v_3 \notin D$ and $w_3$ is $D_{basis}$-forced. Analogously, if $u'$ contacts a third $P_2$ $(w_3,v_3)$ in $N_3$ then $w_3 \notin D$ and $v_3$ is $D_{basis}$-forced.   
\item[$(iii)$] For $u,u',u'' \in N_2$, there is no $4P_2$ $(v_1,w_1)$, $(v_2,w_2)$, $(v_3,w_3)$, $(v_4,w_4)$ in $N_3$ with 
$v_1,v_2,v_3,v_4 \in N(u) \cap N_3$, $w_1,w_2 \in N(u') \cap N_3$, $w_3,w_4 \in N(u'') \cap N_3$.
\end{itemize}
\end{coro}  

\noindent
{\bf Proof.} 
$(i)$: Recall that for every $P_2$ $(v_i,w_i)$ in $N_3$, either $v_i \in D$ or $w_i \in D$. 
If for $u,u' \in N_2$, there is a $3P_2$ $(v_1,w_1)$, $(v_2,w_2)$, $(v_3,w_3)$ with $uv_i \in E$, $u'w_i \in E$, $1 \le i \le 3$, and  
without loss of generality, $v_1 \in D$ and $w_2 \in D$, then by the e.d.s.\ property, $v_3,w_3 \notin D$ and there is no such e.d.s.\ in $G$. 
Thus, there is no $3P_2$ $(v_1,w_1)$, $(v_2,w_2)$, $(v_3,w_3)$ in $(N(u) \cup N(u')) \cap N_3$.

\medskip

\noindent
$(ii)$: Assume that $u,u' \in N_2$ contact a $2P_2$ $(v_1,w_1)$, $(v_2,w_2)$ in $N_3$ with $uv_i \in E$, $u'w_i \in E$, $i \in \{1,2\}$. 
Recall that $v_i \in D$ or $w_i \in D$, $i \in \{1,2\}$, and by the e.d.s.\ property, either $v_1,w_2 \in D$ or $v_2,w_1 \in D$, i.e., $u$ as well as $u'$ have already exactly one $D$-neighbor in $N_3$. 
If $u$ contacts a third $P_2$ $(v_3,w_3)$ in $N_3$, i.e., $uv_3 \in E$ then $v_3 \notin D$ and $w_3$ is $D_{basis}$-forced.
Analogously, if $u'$ contacts a third $P_2$ $(w_3,v_3)$ in $N_3$, i.e., $u'w_3 \in E$ then $w_3 \notin D$ and $v_3$ is $D_{basis}$-forced.   

\medskip

\noindent
$(iii)$: Suppose to the contrary that there is a $4P_2$ $(v_1,w_1)$, $(v_2,w_2)$, $(v_3,w_3)$, $(v_4,w_4)$, with 
$v_1,v_2,v_3,v_4 \in N(u) \cap N_3$, $w_1,w_2 \in N(u') \cap N_3$, $w_3,w_4 \in N(u'') \cap N_3$. 
Recall that for every $P_2$ $(v_i,w_i)$ in $N_3$, $v_i \in D$ or $w_i \in D$, $1 \le i \le 4$, and by the e.d.s.\ property, either $v_1,w_2 \in D$ or $v_2,w_1 \in D$ since $u,u'$ contact a $2P_2$ $(v_1,w_1)$, $(v_2,w_2)$. 

Without loss of generality, assume that $v_1 \in D$ and $w_2 \in D$. But then $v_3,v_4 \notin D$ and $w_3,w_4 \in D$, which is a contradiction by the 
e.d.s.\ property since $u''w_3 \in E$ and $u''w_4 \in E$. Thus, there is no such $4P_2$ $(v_1,w_1)$, $(v_2,w_2)$, $(v_3,w_3)$, $(v_4,w_4)$ in $N_3$ with 
$v_1,v_2,v_3,v_4 \in N(u) \cap N_3$, $w_1,w_2 \in N(u') \cap N_3$, $w_3,w_4 \in N(u'') \cap N_3$.

\medskip

Thus, Corollary \ref{N3no3P2uu'} is shown.
\qed

\medskip

Recall that the e.d.s.\ problem can be done independently for the components in $G[N_2 \cup N_3 \cup N_4]$.
Now let $Q$ be a component in $G[N_2 \cup N_3 \cup N_4]$ with nontrivial components $K_1,\ldots,K_{\ell}$, $\ell \ge 1$, in $G[N_3]$. 
Clearly, in any nontrivial component $K_i$ in $G[N_3]$, there is a $D$-vertex (recall Claim \ref{nontrivialcompinN3} $(iii)$). Then every vertex in $N_2$ has either a $D$-neighbor in a $P_2$ in $N_3$ or in no $P_2$ in $N_3$. 

\medskip

If $\ell = 1$, there is only one nontrivial component $K_1$ in $Q$, and the remaining parts of $Q$ are partial components $Q'$ with independent $V(Q') \cap N_3$. 

\medskip

Recall that there is no $P_4$ in $K_1$ but a $P_2$ in $K_1$, say $(v,w)$ with $v \in D$, and there is a $u \in N_2$ with $uv \in E$. 
Then for the partial components $Q'$ with independent $V(Q') \cap N_3$ and with contact to $u$, the $D$-vertices in $Q'$ are $D_{basis} \cup \{v\}$-forced (recall Claim \ref{v1inDN3v2Dneighbforced}). For the partial components $Q'$ with independent $V(Q') \cap N_3$ and without contact to $u$, it can be done independently.

\medskip

Now assume that $\ell \ge 2$ for component $Q$ in $G[N_2 \cup N_3]$ with nontrivial components $K_1,\ldots,K_{\ell}$. 
Recall that by (\ref{(N2QXjoinY)}), $(V(Q) \cap N^X_2) \join (V(Q) \cap N^Y_2)$. 

\medskip

Recall that by Claim \ref{u123inN2noP7}, there is no $P_7$ $(v_1,u_1,v_2,u_2,v_3,u_3,v_4)$ with $u_1,u_2,u_3 \in N_2$ and $v_1,v_2,v_3,v_4 \in N_3$. 
Thus, for the distance between $K_i$ and $K_j$ in component $Q$, there are at most two such $u_i,u_j \in N_2$ with the same color.
Moreover, recall that for every $P_2$ $(v,w)$ in $V(Q) \cap N_3$, $v$ has only one neighbor $w \in V(Q) \cap N_3$ and $v$ does not contact $N_4$, and analogously,  
$w$ has only one neighbor $v \in V(Q) \cap N_3$ and $w$ does not contact $N_4$.

\begin{clai}\label{P2componinN3}
Let $(x_1,y_1)$ be a $P_2$ in $V(Q) \cap N_3$ and $ux_1 \in E$, $u'y_1 \in E$ with $u,u' \in N_2$. The following statements hold:
\begin{itemize}
\item[$(i)$] If $uv \in E$ with $v \in N_3$ and $w_1,w_2 \in N(v) \cap N_4$ then $x_1 \notin D$ and $y_1$ is $D_{basis}$-forced. 
Analogously, if $u'v' \in E$ with $v' \in N_3$ and $w'_1,w'_2 \in N(v') \cap N_4$ then $y_1 \notin D$ and $x_1$ is $D_{basis}$-forced. 
\item[$(ii)$] If $uv \in E$ with $v \in N_3$, $w_1,w_2 \in N(v) \cap N_4$, and $u'v' \in E$ with $v' \in N_3$, $w'_1,w'_2 \in N(v') \cap N_4$ then there is no such e.d.s.
\end{itemize}
\end{clai}

\noindent
{\bf Proof.} 
$(i)$: Without loss of generality, let $uv \in E$ with $v \in N_3$ and $w_1,w_2 \in N(v) \cap N_4$. 

\medskip

Suppose to the contrary that $x_1 \in D$. Then by the e.d.s.\ property, $v \notin D$, and by Claim \ref{v1inDN3v2Dneighbforced}, $v$ must have only one neighbor 
in $N_3 \cup N_4$, which is a contradiction. Thus, $x_1 \notin D$ and $y_1$ is $D_{basis}$-forced. 
Analogously, if $u'v' \in E$ with $v' \in N_3$ and $w'_1,w'_2 \in N(v') \cap N_4$ then $y_1 \notin D$ and $x_1$ is $D_{basis}$-forced. 

\medskip

\noindent
$(ii)$: Assume that $uv \in E$ with $v \in N_3$, $w_1,w_2 \in N(v) \cap N_4$, and $u'v' \in E$ with $v' \in N_3$, 
$w'_1,w'_2 \in N(v') \cap N_4$. Then by $(i)$, $x_1,y_1 \notin D$ and $x_1,y_1$ do not have any $D$-neighbor but then there is no such e.d.s. 
\qed 

\medskip

If $x_1$ is $D_{basis}$-forced then we can update $D_{basis}:=D_{basis} \cup \{x_1\}$. Analogously, if $y_1$ is $D_{basis}$-forced then we can update $D_{basis}:=D_{basis} \cup \{y_1\}$.

\medskip
  
From now on, assume that for every $P_2$ $(x_1,y_1)$ in $V(Q) \cap N_3$ with $ux_1 \in E$, $u'y_1 \in E$, $u,u' \in N_2$, $uv \in E$, $u'v' \in E$, $v,v' \in N_3$,   $v$ as well as $v'$ have only one neighbor in $N_3 \cup N_4$.  

\medskip

First assume that there is a common neighbor $u \in N_2$ of $P_2$'s $(x_1,y_1)$, $(x_2,y_2)$ in $V(Q) \cap N_3$, say without loss of generality, $ux_1 \in E$ as well as $ux_2 \in E$. If $x_1 \in D$ then $x_2$ is $D_{basis} \cup \{x_1\}$-excluded, and $y_2$ is $D_{basis} \cup \{x_1\}$-forced. 

\medskip

Moreover, if $ux_1 \in E$, $ux_2 \in E$ as well as $u'y_1 \in E$, $u'y_2 \in E$, $u,u' \in N_2$, then $x_1,y_2 \in D$ or $y_1,x_2 \in D$. 
Recall that by Corollary \ref{N3no3P2uu'} $(i)$, there is no such $3P_2$ $(x_1,y_1)$, $(x_2,y_2)$, $(x_3,y_3)$ in $N_3$ with common neighbors $ux_i \in E$, $u'y_i \in E$, $1 \le i \le 3$. 

Moreover, recall that by Corollary \ref{N3no3P2uu'} $(ii)$, if $u$ contacts a third $P_2$ $(x_3,y_3)$ in $V(Q) \cap N_3$ then $x_3 \notin D$ and $y_3$ is $D_{basis}$-forced. Analogously, if $u'$ contacts a third $P_2$ $(x_3,y_3)$ in $V(Q) \cap N_3$ then $y_3 \notin D$ and $x_3$ is $D_{basis}$-forced.

Thus, $D_{basis}$ can be updated, i.e., either $D_{basis}:=D_{basis} \cup \{y_3\}$ or $D_{basis}:=D_{basis} \cup \{x_3\}$, and from no on, assume that there is no such third $P_2$ $(x_3,y_3)$ in $V(Q) \cap N_3$ with $ux_1 \in E$, $ux_2 \in E$ as well as $u'y_1 \in E$, $u'y_2 \in E$. Moreover, assume that for every $2P_2$ 
$(x_1,y_1)$, $(x_2,y_2)$ in $V(Q) \cap N_3$, either $x_1,x_2$ or $y_1,y_2$ do not have any common neighbor in $N_2$.   

\medskip

Now assume that there is no common $N_2$-neighbor between any $P_2$'s in $V(Q) \cap N_3$. For $(x_1,y_1)$, $(x_2,y_2)$ in $V(Q) \cap N_3$, without loss of generality, assume that there are $u_1,u_2 \in N_2$ with $u_1x_1 \in E$ and $u_1x_2 \notin E$ as well as $u_2x_2 \in E$ and $u_2x_1 \notin E$. 
Since by Claim~\ref{u123inN2noP7}, there is no such $P_7$ with three $N_2$-vertices of the same color and four $N_3$-neighbors of the same color, 
$u_1$ and $u_2$ have a common neighbor $v \in N_3$. 

Clearly, $vy_i \notin E$, $i \in \{1,2\}$ since $(x_i,y_i,v)$ does not induce a $P_3$ in $V(Q) \cap N_3$, and   
$v$ does not have any neighbor in $N_3$ (else $u_1$ is a common neighbor between two $P_2$'s in $V(Q) \cap N_3$), i.e., $v$ has exactly one neighbor in $N_4$ (recall Claim \ref{P2componinN3}).   
Now $P=(y_1,x_1,u_1,v,u_2,x_2,y_2)$ induce a $P_7$ in $Q$ with $x_1,y_1,x_2,y_2,v \in V(Q) \cap N_3$ and $u_1,u_2 \in V(Q) \cap N_2$. 

\medskip

First assume that for the $P_7$-midpoint $v \in N_3$ of $P$ in $Q$, $v \in D$: 
Then by the e.d.s.\ property, $x_1,x_2 \notin D$ and thus, $y_1,y_2 \in D$, which are $D_{basis} \cup \{v\}$-forced. 

If there is a next $P_7$ $P'=(x_2,y_2,u'_2,v',u_3,y_3,x_3)$ in $Q$ with $x_2,y_2,x_3,y_3,v' \in V(Q) \cap N_3$, $u'_2,u_3 \in V(Q) \cap N_2$ then by the e.d.s.\ property, $v' \notin D$ since $y_2 \in D$, and a $D$-neighbor $w \in D \cap N_4$ of $v'$ is $D_{basis} \cup \{v\}$-forced. 

Moreover, if $N(u_3) \cap N_3=\{y_3,v'\}$ then, since $v' \notin D$ and $u_3$ must have a $D$-neighbor in $N_3$, $y_3$ is $D_{basis} \cup \{v\}$-forced, and if $|N(u_3) \cap N_3| \ge 3$ then the next subcomponent with $u_3,y_3,x_3$ etc.\ can be done independently.     

\medskip

Now assume that the $P_7$-midpoint $v \notin D$: If $N(u_1) \cap N_3=\{x_1,v\}$ then, since $u_1$ must have a $D$-neighbor in $N_3$, $x_1 \in D$. Analogously, if   
$N(u_2) \cap N_3=\{x_2,v\}$ then, since $u_2$ must have a $D$-neighbor in $N_3$, $x_2 \in D$. 
Now assume that $|N(u_i) \cap N_3| \ge 3$, $i \in \{1,2\}$. 

If $x_1 \in D$ and $|N(u_2) \cap N_3| \ge 3$ then the next subcomponent with $u_2,x_2,y_2$ etc.\ can be done independently.     

If $x_1 \notin D$ and $y_1 \in D$ then, if $N(u_1) \cap N_3=\{x_1,v\}$, $v \in D$, i.e., $v$ is $D_{basis} \cup \{y_1\}$-forced.    
Now assume that $|N(u_1) \cap N_3| \ge 3$, and then the next subcomponent with $u_1,v,u_2,x_2,y_2$ etc.\ can be done independently. 

\begin{coro}\label{EDcomponpoltime}
For every component in $G[N_2 \cup N_3 \cup N_4]$ in this section, there is a polynomial time solution for finding an e.d.s.\ or a contradiction.
\end{coro}

\section{Distance 3 between two $D$-vertices $x,y \in D_{basis}$}\label{xyinDdist3}

\subsection{When there exist black and white $D$-vertices in $N_3$}\label{xyinDdist3blackandwhiteDinN3}

Recall that $x,y \in D_{basis}$ with $dist_G(x,y)=3$, say $(x,y_1,x_1,y)$ induce a $P_4$ in $G$ with $y_1,x_1 \in N_1$, and every $(x,y)$-forced vertex is also in 
$D_{basis}$, i.e., $N_0=D_{basis}$, $N_i$, $i \ge 1$, are the distance levels of $D_{basis}$, and $D \cap (N_1 \cup N_2)=\emptyset$. 
If $N_2$ contains black and white vertices, say $u_1 \in N_2 \cap X$, $u_2 \in N_2 \cap Y$, then there are at least two white and black $D$-neighbors of $u_1,u_2 \in N_2$ in $N_3$, say $v_1 \in D \cap N_3 \cap Y$ with $u_1v_1 \in E$ and $v_2 \in D \cap N_3 \cap X$ with $u_2v_2 \in E$ 
(else there is no such e.d.s.\ with $x,y \in D$). 

\medskip

Recall that by (\ref{vinN3neighbinN3N4}), every vertex in $N_3$ has a neighbor in $N_3 \cup N_4$. 
Assume that for $v_i \in D \cap N_3$, $i \in \{1,2\}$, $(u_i,v_i,w_i)$ induce a $P_3$ with $u_i \in N_2$ and $w_i \in N_3 \cup N_4$. 
Clearly, $w_i \cojoin N_1$, $i \in \{1,2\}$, and since $u_1 \in N_2 \cap X$, $u_2 \in N_2 \cap Y$, we have 
$u_1 \cojoin N(D_{basis} \cap Y)$ as well as $u_2 \cojoin N(D_{basis} \cap X)$. 

\medskip

Clearly, $v_1,v_2$ are not yet $D_{basis}$-forced, but assume that $D_{basis}:=D_{basis} \cup \{v_1,v_2\}$, $N'_0=D_{basis}$, and $N'_i$, $i \ge 1$, are the new distance levels of the new $D_{basis}$. Clearly, again $D \cap (N'_1 \cup N'_2) = \emptyset$. 

\begin{lemma}\label{xyv1v2N2endpointP5}
For every $u \in N'_2$, $u$ is the endpoint of a $P_5$ whose remaining vertices are in $N'_0 \cup N'_1$. 
\end{lemma}

\noindent
{\bf Proof.}
Recall that $v_1,v_2 \in D_{basis}$, $v_1$ is white and $v_2$ is black. Moreover, recall that $(u_i,v_i,w_i)$, $i \in \{1,2\}$, induce a $P_3$ with $u_i \in N_2$ and 
$w_i \in N_3 \cup N_4$ for the previous distance levels $N_2,N_3,N_4$, and then $w_i \cojoin N(D_{basis} \setminus \{v_1,v_2\})$, $i \in \{1,2\}$. 

\medskip

Without loss of generality, assume that $u \in N'_2 \cap X$, i.e., $u$ is black. Then $u$ does not contact $N(v)$ for every $v \in D_{basis} \cap Y$; 
in particular, $u \cojoin (N(y) \cup N(v_1))$. Clearly, $u$ contacts $N(r)$ for some $r \in D_{basis} \cap X$ (possibly $r=x$ or $r=v_2$). 

If $u \cojoin N(x)$ then by Lemma \ref{N2endpointP4} $(i)$, $u$ is the endpoint of a $P_5$ whose remaining vertices are in $N'_0 \cup N'_1$.
Thus assume that $u$ contacts $N(x)$, i.e., $u \in N_2$ for the previous distance level $N_2$.  
 
\medskip

\noindent
{\em Case $1$.} $u_1$ contacts $N(x)$:

\medskip

First assume that $u$ and $u_1$ have a common neighbor $z \in N(x)$. Since $u \in N'_2$ and $v_1 \in D_{basis}$, $uv_1 \notin E$. 
Clearly, $w_1z \notin E$ since in the previous distance levels $N_i$, $i \ge 1$, $z \in N_1$ and $w_1 \in N_3 \cup N_4$. 
Then for $z,u_1,w_1 \in N'_1$ and $v_1 \in N'_0$, $(u,z,u_1,v_1,w_1)$ induce a $P_5$ with endpoint $u$.  

Now assume that $u$ and $u_1$ do not have any common neighbor in $N(x)$, say for $z,z' \in N(x)$, $z \neq z'$, $uz \in E$, $u_1z' \in E$, and 
$uz' \notin E$, $u_1z \notin E$. 
But then $(u,z,x,z',u_1,v_1,w_1)$ induce a $P_7$ with endpoint $u$, which is impossible by Claim \ref{uinN2noendpointP6inN0N1}. 

\medskip

\noindent
{\em Case $2$.} $u_1$ does not contact $N(x)$: 

\medskip 

Since $u_1 \in N_2$ in the previous distance level $N_2$, $u_1$ must have a neighbor in $N_1$, say $u_1s \in E$ with $s \in N(r)$ and $r \in N_0 \cap X$. 
Clearly, $w_1 \cojoin N(r)$, i.e., $w_1s \notin E$ since $w_1 \in N_3 \cup N_4$.

If $u \cojoin N(r)$ then by Lemma \ref{N2endpointP4} $(i)$, $u$ is the endpoint of a $P_5$ whose remaining vertices are in $N'_0 \cup N'_1$. 
Now assume that $u$ contacts $N(r)$. If $u$ and $u_1$ have a common neighbor $s \in N(r)$ then again for $s,u_1,w_1 \in N'_1$ and $v_1 \in N'_0$, 
$(u,s,u_1,v_1,w_1)$ induce a $P_5$ with endpoint $u$. 

Moreover, if $u$ and $u_1$ do not have any common neighbor in $N(r)$, say for $s,s' \in N(r)$, $s \neq s'$, 
$u_1s \in E$, $us' \in E$,  and $us \notin E$, $u_1s' \notin E$, then $(u,s',r,s,u_1,v_1,w_1)$ induce a $P_7$ with endpoint $u$, which is impossible by Claim \ref{uinN2noendpointP6inN0N1}. 

\medskip

Thus, in general, $u$ is the endpoint of a $P_5$ whose remaining vertices are in $N'_0 \cup N'_1$. 

\medskip

Analogously, the same can be shown when $u \in N'_2 \cap Y$, i.e., $u$ is white, and Lemma \ref{xyv1v2N2endpointP5} is shown.
\qed

\medskip

By Lemma \ref{xyv1v2N2endpointP5}, the e.d.s.\ problem in subsection \ref{xyinDdist3blackandwhiteDinN3} can be solved in polynomial time as in Section \ref{N2endpointP5N0N1} when there are black and white $D$-vertices in $N_3$. 

\subsection{When all $D$-vertices in $N_3$ have the same color}\label{xyinDdist3samecolorDinN3}

Recall that by Lemma \ref{N2endpointP4} $(iii)$, $N_6=\emptyset$. 
Without loss of generality, assume that all $D$-vertices in $N_3$ are black, i.e., $D \cap N_3 \subset X$ and $D \cap N_3 \cap Y = \emptyset$. 
Then $N_2 \subset Y$ (else for a black vertex in $N_2$, there is no white $D$-neighbor in $N_3$, and there is no such e.d.s.\ $D$ with $x,y \in D_{basis}$), 
$N_3 \subset X$, $N_4 \subset Y$, and $N_5 \subset X$, i.e., every $N_i$, $2 \le i \le 5$, is independent. Clearly, $N(x)$ is white and thus, $N(x) \cojoin N_2$. 

\subsubsection{When $N_5 \neq \emptyset$}\label{xyinDdist3samecolorDinN3N5nonempty}

If for every $r_5 \in N_5$, $r_5\in D$, i.e., $N_5 \subset D$, then $D_{basis}:=D_{basis} \cup N_5$, and $N'_5=\emptyset$ for the new distance level $N'_5$. 

Now assume that there is an $r_5 \in N_5$ with $r_5 \notin D$. Then $r_5$ must have a $D$-neighbor $r_4 \in D \cap N_4$. Let $(r_0,r_1,r_2,r_3,r_4,r_5)$ induce a 
$P_6$ with $r_i \in N_i$, $0 \le i \le 5$ (possibly $r_0=y$). Then, since by the e.d.s.\ property, $r_3 \notin D$, $r_2$ must have a $D$-neighbor $s_3 \in D \cap N_3$, $s_3 \neq r_3$. 

If $s_3 \cojoin N_4$ then $s_3$ is $D_{basis}$-forced or it leads to a contradiction if $r_2$ has two such neighbors $s_3,s'_3$ with $s_3 \cojoin N_4$, $s'_3 \cojoin N_4$. Now assume that $s_3s_4 \in E$ with $s_4 \in N_4$. 

Then either $(r_2,r_1,r_0,s_3,s_4,r_3,r_4,r_5)$ induce an $S_{2,2,3}$ with midpoint $u_2$ if $s_4r_3 \notin E$ and $s_4r_5 \notin E$, or $s_4r_3 \in E$ or 
$s_4r_5 \in E$; we call it an $S^*_{2,2,3}$.  

Now assume that $D_{basis}:= D_{basis} \cup \{r_0,s_3,r_4\}$ (possibly $r_0=y$). Again, $N'_0=D_{basis}$ and $N'_i$, $i \ge 1$, are the second distance levels of the new $D_{basis}$.

\begin{lemma}\label{r0s3r4inDbasisuinN2endpointP5}
Each vertex in $N'_2$ is the endpoint of a $P_5$ whose remaining vertices are in $N'_0 \cup N'_1$.  
\end{lemma}

\noindent
{\bf Proof.}
Let $u \in N'_2$. Clearly, $u$ has a neighbor $v \in N'_1$. Recall that for $r_0,s_3,r_4 \in N'_0$, $r_0$ is white (possibly $r_0=y$), $s_3$ is black, $r_4$ is white, and $r_1,r_2,r_3,r_5,s_4 \in N'_1$. 

\medskip

\noindent
{\em Case $1$.} $u \in N'_2 \cap X$, i.e., $u$ is black: 

\medskip

Clearly, $u \cojoin N(y)$ since $N(y)$ is black.  
Recall that for the previous distance levels $N_i$, $1 \le i \le 5$, $N_2 \cup N_4 \subset Y$. 
Then $u \in N_3 \cup N_5$ and since $N(x) \subset N_1$, $u \cojoin N(x)$. 
Thus, by Lemma \ref{N2endpointP4} $(i)$, every $u \in N'_2 \cap X$ is the endpoint of a $P_5$ whose remaining vertices are in $N'_0 \cup N'_1$.  

\medskip

\noindent
{\em Case $2$.} $u \in N'_2 \cap Y$, i.e., $u$ is white: 

\medskip

Then $u \cojoin (N(s_3) \cup N(x))$ since $N(s_3) \cup N(x)$ is white. Recall that $u \in N_2 \cup N_4$ for the previous distance levels. 

First assume that $u \in N_4$. Since $N(y) \subset N_1$, $u \cojoin N(y)$, and by Lemma \ref{N2endpointP4} $(i)$, $u$ is the endpoint of a $P_5$ whose remaining vertices are in $N'_0 \cup N'_1$. 

Next assume that $u \in N_2$ and $u$ contacts $N(y)$ (possibly $r_0=y$). Clearly, $ur_5 \notin E$ since $r_5 \in N_5$.  
Since $u \in N'_2 \cap Y$ and $s_3 \in N'_0$, $u \notin N(s_3)$, i.e., $us_3 \notin E$. If $ur_1 \in E$ then $(u,r_1,r_2,s_3,s_4)$ induce a $P_5$ with endpoint $u$. 
Now assume that $ur_1 \notin E$. 

\medskip

If $u \cojoin N(r_4)$ then, since $u$ is white and $N(r_4)$ is black (recall that not only $x,y \in N'_0$ but also $r_0,s_3,r_4 \in N'_0$),  
 and again by Lemma \ref{N2endpointP4} $(i)$, $u$ is the endpoint of a $P_5$ whose remaining vertices are in $N'_0 \cup N'_1$.

\medskip

Now assume that $u$ contacts $N(r_4)$. If $ur_1 \in E$ then $(u,r_1,r_2,s_3,s_4)$ induce a $P_5$ with endpoint $u$. 
Now assume that $ur_1 \notin E$. If $ur_3 \in E$ then $(u,r_3,r_2,r_1,r_0)$ induce a $P_5$ with endpoint $u$. 
Now assume that $ur_3 \notin E$. Recall that $ur_5 \notin E$ since $u \in N_2$ and $r_5 \in N_5$, and let $ur \in E$, $r \in N(r_4)$, $r \neq r_3,r_5$. 

Since by Claim~\ref{uinN2noendpointP6inN0N1}, $(u,r,r_4,r_3,r_2,r_1,r_0)$ does not induce a $P_7$ with endpoint $u$, we have $rr_2 \in E$, 
and then $(u,r,r_2,r_1,r_0)$ induce a $P_5$ with endpoint $u$. 
 
\medskip

Thus, Lemma \ref{r0s3r4inDbasisuinN2endpointP5} is shown.  
\qed

\medskip

Then the e.d.s\ problem can be solved in polynomial time as in Section \ref{N2endpointP5N0N1}. 

\subsubsection{When $N_5 = \emptyset$}\label{xyinDdist3samecolorDinN3N5empty}

Recall that $N_2 \cup N_4$ is white, $N_3$ is black, and now, $N_5=\emptyset$.

\medskip

\noindent
{\bf Case 1.} $|D \cap N_3|=1$, say without loss of generality, $D \cap N_3 \cap X=\{v\}$:

\medskip

Clearly, $v \join N_2$ (else there exists a vertex in $N_2$ which does not have any $D$-neighbor, and there is no such e.d.s.\ $D$ with $x,y \in D_{basis}$), 
and $N_4 \setminus N(v) \subset D$ or it leads to a contradiction by the e.d.s.\ property (else there is no such e.d.s.\ in $G$). 
Then every vertex in $N_3 \setminus \{v\}$ is the endpoint of a $P_5$ whose remaining vertices are in $N_0 \cup N_1 \cup N_2$.  

\medskip

Now we can assume that $v \in D_{basis}$, i.e., $D_{basis}:=D_{basis} \cup \{v\}$, and $N'_i$, $i \ge 1$, are the distance levels of the new $D_{basis}$.  
Then $N'_2=N_3 \setminus \{v\}$ (recall $N_5 = \emptyset$), every $u \in N'_2$ is the endpoint of a $P_5$ whose remaining vertices are in 
$N'_0 \cup N'_1$, and the e.d.s.\ problem in Case 1 with $D \cap N_3=\{v\}$ can be solved in polynomial time as in Section \ref{N2endpointP5N0N1}.

\medskip

Thus, for every possible $v \in D \cap N_3$ with $v \join N_2$, the e.d.s.\ problem can be done in polynomial time. Now assume that no vertex in $N_3$ has a join to $N_2$. 

\medskip

\noindent
{\bf Case 2.} $|D \cap N_3| \ge 2$:

\medskip

Let $v_1,v_2 \in D \cap N_3$ with $P_3$ $(u_i,v_i,w_i)$, and $u_i \in N_2$, $w_i \in N_4$, $i \in \{1,2\}$. Clearly, by the e.d.s.\ property, $(u_i,v_i,w_i)$ and 
$(u_j,v_j,w_j)$, $i \neq j$, induce a $2P_3$ in $G$.
Since $G$ is $P_8$-free bipartite, $u_1$ and $u_2$ must have a common neighbor in $N_1$. 

\medskip

If $D \cap N_4 = \emptyset$ then clearly, $N_3 \subset D$, which leads to a polynomial time solution for an e.d.s.\ or a contradiction.  
Now assume that $D \cap N_4 \neq \emptyset$. If there is only one $D$-vertex $w \in D \cap N_4$ then all vertices in $N_3 \setminus N(w)$ are $D_{basis} \cup \{w\}$-forced, which leads to a polynomial time solution for an e.d.s.\ or a contradiction. 

\medskip

Now assume that $|D \cap N_4| \ge 2$, say $w,w' \in D \cap N_4$, and $vw \in E$, $v'w' \in E$ for $v,v' \in N_3$.    
Thus, for every $w \in N_4$, $N(w) \subseteq N_3$.

Recall that by (\ref{uinN2twoneighbinN3}), for $u_i \in N_2$, $i \in \{1,2\}$, $|N(u_i) \cap N_3| \ge 2$, say $v_i,v'_i \in N(u_i) \cap N_3$ with $v_i \in D$ and $v'_i \notin D$. Then $v'_i$ must have a $D$-neighbor $w'_i \in D \cap N_4$ (possibly $w'_1=w'_2$).  

\medskip

Let $D_{basis}:=D_{basis} \cup \{v_1,w'_1\}$, let $N'_0:=D_{basis}$ and let $N'_i$, $i \ge 1$, be the new distance levels of the new $D_{basis}$. 

\begin{lemma}\label{xyv1v2N2endpointP5Case2.2}
For every $u \in N'_2$, $u$ is the endpoint of a $P_5$ whose remaining vertices are in $N'_0 \cup N'_1$. 
\end{lemma}

\noindent
{\bf Proof.}
Let $u \in N'_2$; clearly, $u$ has a neighbor in $N'_1$. Recall that $x,v_1$ are black and $y,w'_1$ are white. 
Recall that $G$ is connected and by Claim \ref{uinN2noendpointP6inN0N1}, $u$ is no endpoint of any $P_6$ or $P_7$ whose remaining vertices are in 
$N'_0 \cup N'_1$. 

\medskip

\noindent
{\em Case $1$.} $u \in N'_2 \cap X$, i.e., $u$ is black:  

\medskip

Then $u \cojoin (N(y) \cup N(w'_1))$ since $N(y) \cup N(w'_1) \subset N'_1 \cap X$. 
Recall that for the previous distance levels, $N_2 \cup N_4$ is white and $N_3$ is black. Then for black $u \in N'_2 \cap X$, $u \in N_3$, 
and thus, $u \cojoin (N(x) \cup N(y))$. Then by Lemma~\ref{N2endpointP4} $(i)$, $u$ is the endpoint of a $P_5$ whose remaining vertices are in $N'_0 \cup N'_1$. 

\medskip

\noindent
{\em Case $2$.} $u \in N'_2 \cap Y$, i.e., $u$ is white:  

\medskip

Then $u \cojoin (N(x) \cup N(v_1))$ since $N(x) \cup N(v_1) \subset N'_1 \cap Y$. Recall that for $x,y \in N_0$, $N_3$ is black and $N_2 \cup N_4$ is white, 
i.e., $u \notin N_3$ and $u \in N_2 \cup N_4$. 

\medskip

\noindent
{\em Case $2.1$.} $u \in N_4$: 

\medskip

Then again $u \cojoin N(y)$, and by Lemma~\ref{N2endpointP4} $(i)$, $u$ is the endpoint of a $P_5$ whose remaining vertices are in $N'_0 \cup N'_1$. 

\medskip

\noindent
{\em Case $2.2$.} $u \in N_2$:

\medskip

If $u \cojoin N(y)$ then by Lemma \ref{N2endpointP4} $(i)$, $u$ is the endpoint of a $P_5$ whose remaining vertices are in $N'_0 \cup N'_1$. Now assume that $u$ contacts $N(y)$, say $ux' \in E$ with $x' \in N(y)$ (possibly $x'=x_1$). 

Since $u_1 \in N_2$, $u_1$ must have a neighbor in $N_1$. First assume that $u_1$ contacts $N(y)$. 
Since $u \in N'_2$ and $v_1 \in N'_0$, $u \notin N(v_1)$, i.e., $uv_1 \notin E$. 

Clearly, $w_1 \cojoin N(y)$ since $w_1 \in N_4$. If $u$ and $u_1$ have a common neighbor $x' \in N(y)$ then $(u,x',u_1,v_1,w_1)$ induce a $P_5$ with endpoint $u$. Now assume that $u$ and $u_1$ do not have any common neighbor in $N(y)$, say $ux' \in E$, $u_1x'' \in E$, with $x',x'' \in N(y)$, $x' \neq x''$. But then $(u,x',y,x'',u_1,v_1,w_1)$ induce a $P_7$ with endpoint $u$, which is impossible by Claim~\ref{uinN2noendpointP6inN0N1}. 

\medskip

Now assume that $u_1$ does not contact $N(y)$, i.e., $u_1 \cojoin N(y)$. Since $u_1 \in N_2$, $u_1$ must have a neighbor $r \in N_1$. Then $ry \notin E$ and 
$r$ must have a $D$-neighbor  $s \in N_0$, i.e., $rs \in E$. If $u \cojoin N(s)$ then by Lemma \ref{N2endpointP4} $(i)$, $u$ is the endpoint of a $P_5$ whose remaining vertices are in $N'_0 \cup N'_1$. Thus assume that $u$ contacts $N(s)$. Clearly, $w_1 \cojoin N(s)$ since $w_1 \in N_4$. If $u$ and $u_1$ have a common neighbor $r \in N(s)$ then $(u,r,u_1,v_1,w_1)$ induce a $P_5$ with endpoint $u$. Now assume that $u$ and $u_1$ do not have any common neighbor in $N(s)$, say $ur' \in E$, $u_1r \in E$, with $r,r' \in N(s)$, $r \neq r'$, $ur \notin E$, $u_1r' \notin E$.  But then $(u,r',s,r,u_1,v_1,w_1)$ induce a $P_7$ with endpoint $u$, which is impossible by Claim~\ref{uinN2noendpointP6inN0N1}. 
      
\medskip

Thus, $u$ is the endpoint of a $P_5$ whose remaining vertices are in $N'_0 \cup N'_1$, 
and Lemma \ref{xyv1v2N2endpointP5Case2.2} is shown.
\qed

\medskip

Then the e.d.s\ problem can be done in polynomial time as in Section \ref{N2endpointP5N0N1}. 

\medskip

Finally, Theorem \ref{EDP8frbippol} is shown.

\section{Conclusion}

{\bf Open problem}: What is the complexity of ED for $P_k$-free bipartite graphs, $k \ge 9$?

\medskip

\noindent
{\bf Acknowledgment.} The second author would like to witness that he just tries to pray a lot and is not able to do anything without that - ad laudem Domini.

\begin{footnotesize}

\end{footnotesize}

\end{document}